\documentclass[12pt,preprint]{aastex}

\usepackage{graphicx,subfigure}
\usepackage{amsmath}
\usepackage{rotating}
\usepackage{epsfig}
\usepackage{epstopdf}

\newcommand\bld[1]{\mbox{\boldmath $#1$}}

\newcommand{\bnabla}{\bld{\nabla}}

\newcommand{\amr}{{\tt LA-COMPASS }}

\newcommand{\del}{\partial}

\newcommand\pc{{\rm\,pc}}

\newcommand\cms{{\rm\,cm\,s^{-1}}}
\newcommand\gcm{{\rm\,g\,cm^{-3}}}
\newcommand\erg{{\rm\,erg}}
\newcommand\ergs{{\rm\,ergs^{-1}}}

\newcommand\msun{{\rm\,M_\odot}}

\newcommand{\be}{\begin{equation}}
\newcommand{\ee}{\end{equation}}
\newcommand{\bea}{\begin{eqnarray}}
\newcommand{\eea}{\end{eqnarray}}

\shortauthors{Guan, Li, and Li}
\shorttitle{}
\begin{document}

\title{Relativistic MHD Simulations of Poynting Flux-Driven Jets}
\author{Xiaoyue Guan \altaffilmark{1}, Hui Li \altaffilmark{1},
 and Shengtai Li \altaffilmark{1}}

\altaffiltext{1}{Theoretical Division, Los Alamos National Laboratory, Los Alamos, 
NM; guan@lanl.gov}

\begin{abstract}

Relativistic, magnetized jets are observed to propagate to very large distances in 
many Active Galactic Nuclei (AGN).  We use 3D relativistic MHD (RMHD) simulations to 
study the propagation of Poynting flux-driven jets in AGN.  These jets are assumed 
already being launched from the vicinity ($\sim 10^3 $ gravitational radii) of 
supermassive black holes. Jet injections are characterized by a model described in 
\cite{li06} and we follow the propagation of these jets to $\sim$ parsec scales.  
We find that these current-carrying jets are always collimated and 
mildly relativistic. When $\alpha$, the ratio of toroidal-to-poloidal magnetic flux injection, 
is large the jet is subject to  non-axisymmetric current-driven instabilities (CDI) 
which lead to substantial dissipation and reduced jet speed.  However,  
even with the presence of instabilities, the jet is not disrupted and will continue to propagate 
to large distances. We suggest that the relatively weak impact by the instability is
due to the nature of the instability being convective and the fact that the 
jet magnetic fields are rapidly evolving on Alfv\'enic timescale. We present the detailed 
jet properties and show that far from the jet launching region, a substantial amount of 
magnetic energy has been transformed into kinetic energy and thermal energy, 
producing a jet magnetization number $\sigma < 1$. In addition, we have also 
studied the effects of a gas pressure supported ``disk'' surrounding the injection region
and qualitatively similar global jet behaviors were observed.  We stress that jet collimation, 
CDIs, and the subsequent energy transitions are intrinsic features of current-carrying jets.

\end{abstract}

\section{Introduction}

Relativistic jets, such as the famous kpc jet in M87, are observed in many 
active galactic nuclei (AGN) systems through multi-wavelength observations. 
AGN jets are collimated, magnetized, mildly relativistic ($\gamma \sim 10$), 
and can travel to large distances (kpc or even Mpc scales). Peculiar spatial 
structures such as knots  are often observed in various locations along the 
direction of jet propagation (e.g. \cite{biretta91}). Monitoring of jet radiation has 
also revealed a range of jet time variabilities (minutes to years), including recently 
observed TeV flares with a variability timescale of minutes (e.g. \cite{aharonian07, albert07}), 
although the mechanisms that are responsible for variabilities are under debate. 
There are still many unresolved problems associated with relativistic jets, such as jet 
composition ($\rm {e^{+}/e^{-}}$pairs vs. $\rm {e^{-}/p^{+}}$ plasma), jet stability, 
particle acceleration/deceleration mechanisms, and jet emission mechanism. 

It is widely accepted that relativistic jets in AGN systems are powered through 
some magnetic processes, and the most likely mechanism is the so-called 
Blandford-Znajek process (Blandford \& Znajek 1977, B-Z hereafter),  where the 
primary energy source is the spin of black hole but transferred via magnetic fields.   
In recent years, development in 
numerical general relativistic magnetohydrodynamics (GRMHD) and force-free 
electrodynamics (FFEM) techniques 
(e.g. \cite{komi99, mg04,devilliers03,devilliers05,mc05,beckwith08,mb09}) has enabled 
time-dependent studies
of the formation and evolution of relativistic jets, sometimes in connection 
with the detailed accretion processes. Moreover, it has been shown numerically that the
B-Z mechanism is capable of powering a magnetically dominated jet with a 
relativistic Lorentz factor up to $\gamma \sim 10$.  In some accretion-type 
simulations such as  \cite{mb09}, although current-driven instabilities (CDI) 
with a $m=1$ kink mode are observed,  jet can get collimated and propagate to 
$\sim 10^3 GM/c^2$, where $GM/c^2$ is the gravitational radii of the black hole,  
without being disrupted nor having much dissipation. These first-principle simulations
have the advantages of exploring the important dynamics of accretion together with
magnetized jet formation. However, due to the 
extreme numerical requirements to resolve the accretion disk dynamics, 
it is very difficult to examine how these jets will evolve beyond several 
thousands of gravitational radii and over astronomically significant timescales. 
Furthermore, observations of jets down to several thousand gravitational radii of the black hole
have been very difficult to obtain, making comparisons between theory/simulations
and observations challenging. 

Another class of jet models is  focused more on the detailed properties of jets in 
their propagation process after they are launched 
\citep{lba00, bk02, nm04, oneill05, li06, naka06, naka07, naka08, komi07, moll08, 
mignone10,mizuno09, mizuno11, oneill12}. They typically adopt an 
MHD or relativistic MHD (RMHD) approach, utilizing 
some boundary conditions to represent a jet injection, and following the jet propagation. 
Simulations of these models can be either on relatively smaller scales, which are focussed 
on the local properties of the flow, or on relatively large scales ($\sim$ kpc), 
where the jet interacts with the surrounding intergalactic medium.  
When a high-velocity, magnetized jet travels through its environment, it could be
subject to instabilities such as magnetic Kelvin-Helmholtz instability due to 
the shear (e.g., see discussions in \cite{bk02,hardee07}),  and/or current-driving 
instabilities when there are strong toroidal fields and/or rotation 
(e.g., see discussions in \cite{mizuno09,narayan09}).  However, the 
long-term consequences of these 
instabilities and how the properties of the localized jet can be transformed into 
observed jet features are not clear.  One particular focus of this type of research is 
to identify the energy transition mechanism (sometimes called the 
jet $\sigma$ problem; $\sigma$ is the jet magnetization parameter; 
see \cite{rg74}) which transforms a magnetically dominated jet deep in 
the gravitational potential of the black hole to possibly kinetically dominated jet 
on larger scales (e.g., as discussed in \cite{lind89} for FR II jets $\sigma \ll 1$). 
\cite{begelman98} has suggested that current-driven instabilities 
can be used to tackle the energy transition problem, and numerical 
simulations by \cite{mizuno09}, \cite{oneill12} have shown CDIs can 
indeed transform jet magnetic energy into kinetic energy. 

Here we present new simulations of magnetic flux-driven relativistic AGN jet 
using RMHD code \amr(\textbf{L}os \textbf{A}lamos \textbf{COMP}utational 
\textbf{AS}trophysics \textbf{S}uite). Assuming that a Poynting-flux dominated jet 
can steadily propagate to $\sim 10^3$ gravitational radii as suggested by 
current generation of GRMHD black hole accretion simulations,  
we adopt the approach of using an injection region with a size $\sim 10^3$ 
gravitational radii and follow the jet evolution out to tens/hundreds pc scales. 
The injected magnetic field has a geometry of ``closed" field lines that are confined
in spatial extent, different from the 
classic split monopole configuration which has an unconfined flow (see discussions in 
\cite{komi09, tch09}). To our knowledge, this is the first time that a RMHD jet 
can be followed to this observation scale.  This paper is also the first of a 
series of papers studying relativistic jets properties.   

The paper is organized as follows. In \S 2 we give a brief description of the RMHD 
code and how the injection is implemented in our models. In \S 3 we present a fiducial model 
where we analyze the properties of the simulated jets in detail, including jet morphologies, 
energetics, and instabilities. We then describe how these properties depend on model parameters 
such as the injected field geometry, disk confinement, and resolution. 
A summary and discussions are given in \S 4.

\section{Numerical Methods and Model Set-up}
\subsection{RMHD Code }

We use a 3D RMHD code based on evolving fluid equations using 
higher-order Godunov-type finite-volume methods. The ideal MHD code is part of 
the code \amr, which was first developed at Los Alamos National Laboratory \citep{ll03} 
and has been used on a range of astrophysical MHD simulations, including the 
jet collimation and stability problems.

The set of relativistic MHD equations can be written in the following conservative form, 
\begin{equation}
\del_{t}\bld{U}  +\del_{i}\bld{F}^{i}  = \bld{S},
\end{equation}
where $i$ denotes a spatial index. First, a set of conserved variables $\bld{U} = (D, M_x, M_y,M_z,B_x,B_y,B_z,E)^{\rm T}$ is
\begin{equation}
{\bld{U}} \equiv  \left({
        \begin{matrix}
        \rho \gamma \\
         (\rho h \gamma ^2 + \bld{B}^2)v_x - (\bld{v}\cdot \bld{B})B_x \\
         (\rho h \gamma ^2 + \bld{B}^2)v_y - (\bld{v}\cdot \bld{B})B_y \\
         (\rho h \gamma ^2 + \bld{B}^2)v_z - (\bld{v}\cdot \bld{B})B_z \\
         B_x \\
         B_y \\
         B_z \\
         \rho h \gamma ^2 -p +\frac{\bld{B}^2}{2} +\frac{\bld{v}^2\bld{B}^2}{2}-\frac{\bld{v}\cdot\bld{B}}{2}
           \end{matrix}}
          \right),
\end{equation}
where $v_i$ and ${\bf B}^{i}$ are the usual velocity and  magnetic field three-vector, and $\gamma$ is the Lorentz factor $\gamma = (1-v^2/c^2)^{-1/2}$.

Second,  a set of fluxes $\bld{F}^i$, where the flux in the x-direction, is given as
\begin{equation}
{\bld{F}^{x}} \equiv \left({
  \begin{matrix}
      D v_x \\
       M_x v_x -\gamma^{-1}b_x B_x + p\\
       M_y v_x - \gamma^{-1}b_y B_x\\
       M_z v_x - \gamma^{-1}b_z B_x\\
       0\\
         B_y v_x - B_x v_y\\
       B_z v_x - B_x y_z\\
       M_x
  \end{matrix}}  
\right),
\end{equation}
where $b_{i}$ are the usual magnetic field four-vector.

Third,  a set of source is
\begin{equation}
\bld{S}  = (\dot D, \dot{M_x}, \dot{M_y} , \dot{M_z}, \dot{B_x}, \dot{B_y}, \dot{B_z}, \dot{E})^{\rm T}, 
\end{equation}
where $h = 1 + \Gamma p/[(\Gamma -1) \rho]$ is the specific enthalpy, and $\Gamma$ is the adiabatic index.  To solve the approximate Riemann problem, we use the HLL flux with parabolic piece wise reconstruction method by \cite{cw84}. 
Note for RMHD code, the set of primitive variables used  for interpolation are
\begin{equation}
{\bld {P}} \equiv (\rho,  v^i, B^i, u )^{\rm T},
\end{equation}
and they are recovered from conservative variables from an iterative algorithm where Newton-Raphson method is implemented. 

Together with no-monopole constrain 
\begin{equation}
  \partial_{i}{{\bf B}^i} = 0, \label{xg_non_mono}
\end{equation}
 and a
description of thermal dynamics the equation system is complete. Numerically, we use a staggered mesh for magnetic fields, and use Constrained-Transport (CT) method to evolve induction equations.  

In the models we use an ideal gas equation of state (EOS),
\begin{equation} 
p=(\Gamma - 1)u, 
\end{equation}
where $u$ is the internal energy density. In this work we use $\Gamma = 5/3$. We have found that using a relativistic EOS with $\Gamma = 4/3$ gives very similar results for the jet properties studied in the work. 

Because the code conserves total energy and there is no explicit cooling, all the 
heat generated by the dissipation (both physical and numerical) in the jet propagation 
process will be captured by the code (see detailed discussion in \S 3). For the jet problem, 
in the total energy equation we have adopted the common practice to exclude the rest 
mass energy from the total energy and the corresponding energy flux. This is because in 
the vast region where total energy is dominated by the rest mass energy, when we need to 
get the other energetics, the subtraction of a large number from the other one may not be accurate.   

\subsection{Our Model}

The basic framework of our 3D simulations involves two key parts: First, the initiation 
of the jet is through a (continued) injection process within a small volume of size $r_{\rm inj}$. 
This is supposed to mimic the outcome of accretion on the supermassive black hole plus
the magnetized jet formation.
Second, the Lorentz force of the injected magnetic fields (and mass) will cause the magnetic fields
to expand into a pre-existing low density, low 
pressure and unmagnetized background plasma with a size that is several hundred times
larger than $r_{\rm inj}$ in all directions. This is supposed to mimic the propagation of 
relativistic jet through the interstellar medium near the galaxy center on $\sim$ tens of pc scales.

With this approach, the critical questions we hope to address include: 1) whether the jet will be 
collimated on scales much larger than $r_{\rm inj}$; 2) whether the jet will be stable;  
and 3) how efficient the energy conversion 
processes inside the jet will be.  Ultimately, these results could contribute to, among other things, 
understanding both the observed jet structures on those scales and physical conditions 
for multi-wavelength jet emissions.

\subsection{Injection of Magnetic Field and Mass}

In order to drive an injection, we have implemented source terms in 
the RMHD equations at each time-step, 
similar to the method used in \cite{li06}. The injected magnetic flux 
has both a poloidal and toroidal component. In cylindrical coordinates $(r, \phi, z)$ 
the poloidal flux function is axisymmetric and
has a form of 
\begin{equation}
\Phi(r,z) = B_{{\rm inj},0} r^2\exp(-\frac{r^2+z^2}{r^2_{{\rm inj},B}}),
\end{equation}
which relates to the $\phi$ component of vector potential $A_{\phi}$ with 
$\Phi(r,z) = r A_{\phi}$. From $\Phi(r,z)$ one can calculate the poloidal field injection functions 
\begin{equation}
B_{{\rm inj},r} = -\frac{1}{r}\frac{\del \Phi}{\del z} = 2B_{{\rm inj},0}\frac{zr}{r^2_{{\rm inj},B}} \exp(-\frac{r^2+z^2}{r^2_{{\rm inj},B}}),\label{Binjeqn1}
\end{equation}
and
\begin{equation}
B_{{\rm inj},z} = \frac{1}{r}\frac{\del \Phi}{\del r} = 2B_{{\rm inj},0}(1-\frac{r^2}{r^2_{{\rm inj},B}})\exp(-\frac{r^2+z^2}{r^2_{{\rm inj},B}}),\label{Binjeqn3}
\end{equation}
where $B_{{\rm inj},0}$ is a normalization constant for field strength and 
$r_{{\rm inj},B}$ is the characteristic radius of magnetic flux injection. 
This form of magnetic fields contains closed poloidal field lines, which causes 
$B_z$ to change directions beyond $r_{{\rm inj}, B}$ with no net poloidal flux. 

The toroidal field injection function is 
\begin{equation}
B_{{\rm inj},\phi}  = \frac{\alpha\Phi}{r} = B_{{\rm inj},0} \alpha r\exp(-\frac{r^2+z^2}{r^2_{{\rm inj},B}})~.
\end{equation}
Here $\alpha$ is a constant parameter and it has the unit of inverse length scale.  
This parameter specifies the ratio of toroidal to poloidal flux injection rate.  
As demonstrated in \cite{li06}, the poloidal and toroidal fluxes are roughly equal 
when $\alpha \sim 2.6$. In our simulations, we typically use $\alpha >> 1$.
The assumption here is that the rotation of the black hole at the base of jet launching 
location will wind up the poloidal field through the B-Z effect and introduce a large toroidal component. 
The injected magnetic fields are given as
\begin{equation}
\dot{B}_{\rm inj} = \gamma _b {\bld B}_{\rm inj},
\end{equation}
where $\gamma_b$ is the characteristic rate of magnetic injection.  
In all our numerical models $\gamma_b$ is set to a constant so that the magnetic energy
injection rate is roughly constant as well.\footnote{Constant injection of magnetic fields 
over a region of $r_{{\rm inj}, B}$ can be
inherently acausal.  However, since our simulations extend in spatial scales $\gg r_{{\rm inj}, B}$ and
in temporal scales $\gg r_{{\rm inj}, B}/c$, the causality concern is somewhat limited.} 

Our numerical model also has mass injection in the injection region.  There are two 
motivations to consider mass flux injection: the first is that it is possible that matter can 
enter the jet at its launching location, although the details of the mass loading is unknown; 
the second motivation is to maintain a certain density floor in the computational domain as the 
magnetic dominated flow expansion tends to introduce extremely low density region.  
The rest mass density injection function is
\begin{equation}
\dot{\rho}_{\rm inj} = \gamma_{\rho} \rho_0 \exp(-\frac{r^2+z^2}{r_{{\rm inj},\rho} ^2}),
\end{equation}
where $r_{{\rm inj},\rho}$ and $\gamma_{\rho}$ are the characteristic radius and rate of mass injection.  

Our numerical models also allows a jet velocity injection in the $z$ direction, and the $v_z$ 
injection function at the central region is
\begin{equation}
v_{{\rm inj},z} = v_{{\rm inj}, 0} \frac{z}{r_{{\rm inj},\rho}} \exp(-\frac{r^2+z^2}{r_{{\rm inj},\rho} ^2}),
\end{equation} 
where $v_{{\rm inj},0}$ is the characteristic velocity, which is often taken as $0.5c$. It turns out that both the total injected mass and total injected kinetic energy are small so they do not affect the overall jet dynamics.

Notice that for simplicity we have chosen not to include initial plasma rotation in our injection scheme.  
Rotation is certainly a factor to consider in jet models, and it has been argued to be important in 
stabilizing jet (e.g. \cite{tomi01, mb09}).  However, it is not clear whether rotation will play a significant 
role on the scales our models correspond to, therefore we do not include rotation in the initial 
conditions and just focus on the limit when the rotation is small. 
The $\phi$ component of the Lorentz force $({\bld J}\times{\bld B})_{{\rm inj},\phi}$ resulted 
from the injected magnetic flux is zero, the evolution of the total magnetic flux, however, could still 
introduce rotation to the gas.  From the models we indeed find 
that the rotation effect is small (see discussion in \S 4.)

Numerically, we treat injection as a source step at the end of each time step.  
For RMHD, the most straightforward way of injection is to add source terms directly 
to the updated primary variables ${B^i}$ and $\rho$, and add an injected momentum 
source to the updated $z$ momentum, as $v_{{\rm inj}, z}$ only applied to the injected 
mass at each step.  Our code is formulated to conserve the total energy. Since the injection 
step will increase total energy at each time step,  we calculate the new total energy at 
the end of each injection step. 

For the injection scheme,  again for simplicity, we have chosen $ r_{{\rm inj},\rho} = r_{{\rm inj}, B}  = r_{\rm inj}$, 
therefore both matter and fields injection are confined within $r_{\rm inj}$. 
The form of magnetic field injection functions guaranties the divergence free nature 
of the injection field. We have also observed $\bnabla \cdot {\bf B} < 10^{-8}$ throughout the simulation 
in all the computational domain.  The mass injection rate is set to be very small to satisfy 
the plasma thermal $\beta \ll 1$ and plasma $\sigma=B^2/(4\pi \gamma ^2 \rho c^2) \gg 1$. 

We adopt a uniform Cartesian $(x,y,z)$ grid with a size of $x = [-L_x/2,L_x/2], y =[-L_y/2, L_y/2], 
z= [-L_z/2, L_z/2]$.  Outflow boundary conditions are enforced on the primary variables.  
The initial grid is filled with a uniform plasma background with a finite gas density 
$\rho_0$ and pressure $P_0$. The initial magnetic field structure has the same form 
as the magnetic injection function Eqn(\ref{Binjeqn1}-\ref{Binjeqn3}) with a strength 
normalization $B_0$.  The injection region is located at the origin of the box with 
an injection radius of $r_{{\rm inj}}$.

In all models we choose $\rho _ 0 =1, P_0 = 10^{-5}, r_{\rm inj} =1, c = 1$. Other units of 
physical quantities for normalization are listed in Table \ref{units}.  To put these numbers in 
an astrophysical context, assuming a background number density of $10^2 ~{\rm cm}^{-3}$ 
and background temperature of $5~ {\rm keV}$, the code sound speed is  $c_s =0.0041c$ 
which corresponds to a physical sound speed of $8.93\times 10^7 {\rm cm s}^{-1}$.   
The code magnetic strength $B_0 = 1$ corresponds to a physical magnetic field of 
$1.38{\rm G}$ and a physical Alfv\'{e}n speed of $ v_{{\rm A},0} \sim 0.707c$. 
Note in all our models we have initial $ c_s \ll v_{{\rm A}, 0} < c $. For the code length scale,  
we choose  injection region size $r_{{\rm inj}} =1$,  and if this corresponds to $1000GM/c^2$, 
then for a supermassive black hole like M87($M_{\rm BH} = 3\times 10^9 \msun$), 
the injection region has a physical size of  $\sim 0.143$pc.  Our computational domain 
usually has a size of $10^2 - 10^3 r_{\rm inj}$, and this corresponds to a physical domain 
size of $14.3 - 143$ pc.  In the code, $t = 1$ then equals to the light crossing time scale for 
the injection region, and it corresponds to a physical time scale of 
$0.47$yr.   We usually follow the jet propagation for a few hundreds to thousands of years.

\section{Results: Relativistic Jet Propagation}

In this work we follow the propagation of relativistic magnetic-flux driven jets from 
$\sim 10^3$ gravitational radii to tens of pc scales where they are often observed. We are 
particularly interested in the jet morphology, whether current-driven instability will occur 
along the way, and if it does, how these instabilities will affect the jet properties. 
Here we first present a fiducial model to give the detailed accounts of the jet propagation.

\subsection{Fiducial Model}

In our fiducial model we have $\alpha = 10$ for magnetic injection.  
The injection rate is $\gamma _\rho = \gamma_b = 1$.  The initial magnetic field strength 
is $B_0 = 0.3$, the magnetic field injection coefficient is $B_{{\rm inj},0} = 0.2 $.  
The jet velocity injection coefficient is $v_{{\rm inj},0} = 0.5$. The computational 
grid has a size of $L_x = L_y = 150, L_z = 400$ with a resolution $N_x =  N_y = 300, 
N_z = 800$. We run the simulation to $t_f = 1500$. 

\subsubsection{Jet Properties}

\begin{sidewaysfigure}
\centering
\scalebox{0.9}{\includegraphics{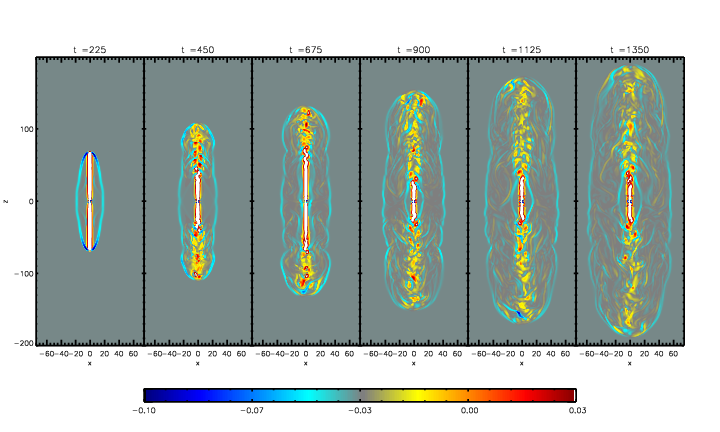}}
\caption{Snapshots of $j_z$ for the fiducial model showing the evolution of
jet propagation and expansion. These snapshots are taken at 
the $y=0$ plane of a 3D simulation, at 
$t = 225, 450, 675, 900, 1125, 1350$, respectively.  
The spatial scales are normalized by $r_{\rm inj}$.
Magnetic energy and flux are injected at the origin $x=y=z=0$
within $r = 1$. The magnetic structure expands 
to form both a collimated jet along the $z-$axis that
carries a strong (positive) outgoing current 
(indicated by the white-red-yellow
color) and a ``cocoon'' enclosing the jet structure with the 
(negative) return current (blue color). The jet continues to propagate
despite becoming unstable. With the instability, both the outgoing and
return current paths show complicated structures, although the overall
outgoing and return current patterns remain. 
}
\label{fig:fidjzcompo}
\end{sidewaysfigure}

\begin{sidewaysfigure}
\centering
\scalebox{0.97}{\includegraphics[width=0.2\textwidth]{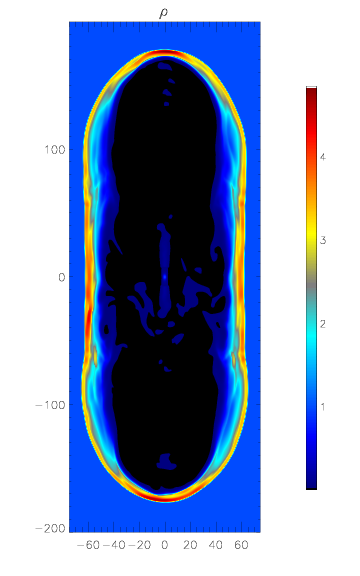}} 
\scalebox{0.97}{\includegraphics[width=0.2\textwidth]{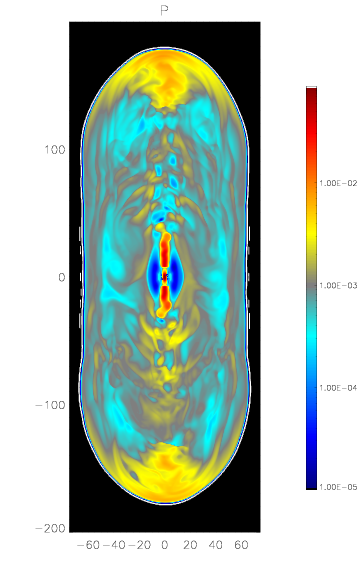}} 
\scalebox{0.97}{\includegraphics[width=0.2\textwidth]{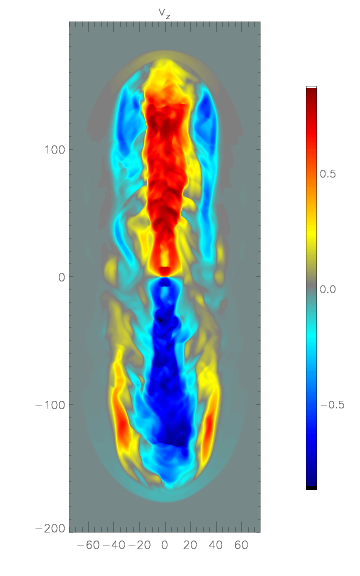}} 
\scalebox{0.97}{\includegraphics[width=0.2\textwidth]{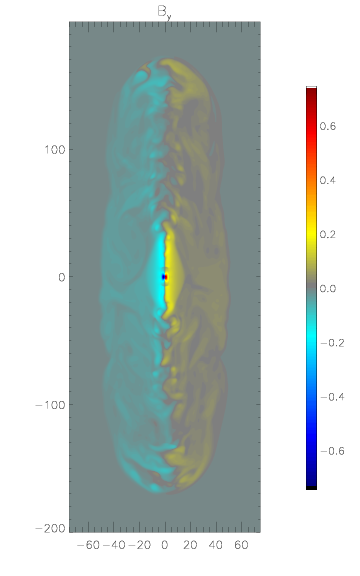}}
\scalebox{0.97}{\includegraphics[width=0.2\textwidth]{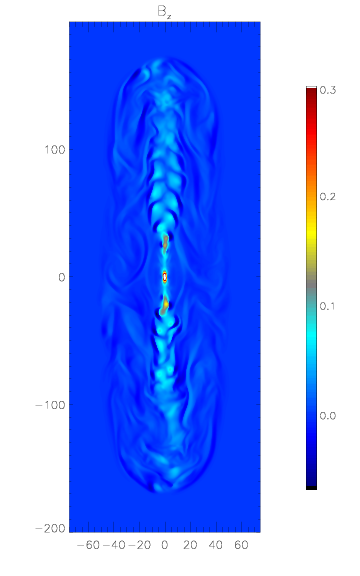}} 

\caption{Snapshots of $\rho, P, v_z, B_y, B_z$ at a relatively late time $t = 1125$
and at the $y = 0$ plane for the fiducial run. The expansion is obviously 
much faster along the vertical direction than that in the transverse direction.
A strong hydrodynamic shock is formed all around the (mildly) relativistically
expanding outer boundary. The jet velocity is relativistic
along the $z-$axis with $\gamma \sim$ a few but slows down significantly 
near the jet fronts. Magnetic fields fill up the volume enclosed by the swept-up 
hydrodynamic shell. The poloidal field dominates along
the $z-$ axis but toroidal field dominates elsewhere. 
}
\label{fig:fidvars}
\end{sidewaysfigure}

Figures \ref{fig:fidjzcompo} and \ref{fig:fidvars} show the overall morphology and evolution of
the jet propagation. Over scales that are much larger than $r_{\rm inj}$, we find that the 
magnetic fields form an elongated structure that stays highly collimated, with the central axis
(along $z$) having roughly a cylindrical shape without an obvious opening angle. While the
central axis of the jet undergoes instabilities, the overall collimation and propagation still remain
(to as long as we have simulated). 
The magnetic structure is enclosed by a hydrodynamic structure
that consists mostly of a strong shock that is propagating into the background and sweeping
up the material into a shell. 

Figure \ref{fig:fidjzcompo} shows several snapshots of the z-component of current density 
$j_z$ at the $y=0$ plane. Because the injected fields possess a dipole-like poloidal field 
structure plus a toroidal field proportional to the flux function, the $j_z$ distribution has the 
overall structure that it contains an ``outgoing'' (positive) current along the central axis and 
a ``return'' current mostly in a thin shell encasing the structure. The location of the return current
separates the magnetized interior from the non-magnetized outer region.  
Before $t \leq 225$, the jet appears to be 
propagating with little signature of nonlinear instabilities, while around $t \sim 300$,  
significant nonlinear instabilities first start to appear at the jet front, indicated by
small wiggles with characteristic length scale $\sim$ a couple of tens $r_{\rm inj}$.  
At the late time many filamentary structures start to appear in jet front as the jet propagates 
further, while the central high $j_z$ region keeps almost the same vertical extent.  
It is interesting to note that the return current has maintained a quite axisymmetric cocoon-like shape
throughout the duration of the run. 
At the late time, along the axis, the $j_z$ distribution splits into two parts: further away from the injection, 
the $j_z$ current density becomes highly unstable; whereas closer to the injection region 
with $|l_z| \leq 50~ r_{\rm inj}$,
it stays quasi-stable with relatively high peak current values (up to $j_{z,{\rm max}} = 3.2$,
not shown in the figure), presumably due to the strong injection.

To illustrate the jet properties at the late time, in Figure \ref{fig:fidvars}, 
we plot the snapshots of gas density $\rho$, gas pressure $P$, $z-$component of 
the gas three-velocity $v_z$,  $y-$component of magnetic field $B_y$, 
and $z-$component of magnetic field $B_z$.  
In the density plot, there is a very thin layer of gas at the shock front with a maximum density  
$\rho _{\rm max} = 4.8$,  while inside this shell there is an extremely low density region with 
a minimum density $\rho_{\rm min} = 7.9 \times 10^{-4}$. This is a result of 
most of the uniform background gas being pushed away by the magnetic-dominated jet as it 
expands into the environment.  Note in the inner $|l_z| \leq 40$ region there's a small amount of 
gas which follows where the strong current is. This is because we inject a small amount of 
gas into the computational domain.  At the end of this simulation, we have injected a total 
of $M_{\rm inj} = 8688 \rho_0 r_{\rm inj}^3$, which for the parameters we specified at the 
end of \S 2, corresponds to a total mass injection of $1.3\times 10^{35} ~{\rm g}$ and 
a mass injection rate of $0.09\msun{\rm /yr}$. 
For the gas pressure, it is evident that the shock front has a higher pressure in the $z$ 
direction than in the horizontal direction, presumably due to the stronger expansion along the $z$ direction.
For the gas velocity, we get the maximum Lorentz factor of the plasma flow 
is $\gamma_{\rm max} \sim 2.7$ 
and the maximum Lorentz factor generally increases with time during the run.   
For $v_z$, we see that while around the $r=0$ axis the gas is mostly moving outward, there's also 
a returning component at larger $r$ due to the magnetic field structure we have used in our model. 
For the $B_y$ and $B_z$ plots, they show that, along the radial direction, 
the jet has a magnetic dominated core with $B_z$ being dominant at $r=0$ but $B_y$ 
becomes dominant at large $r$. Along the $z$ direction, there is a magnetic dominated region
with $|l_z| \leq 50 r_{\rm inj}$ that is followed by a more smoothly decreasing region out to the
vertical extent of the jet. Overall, a magnetized central spine is always present. 

\begin{figure}
\plotone{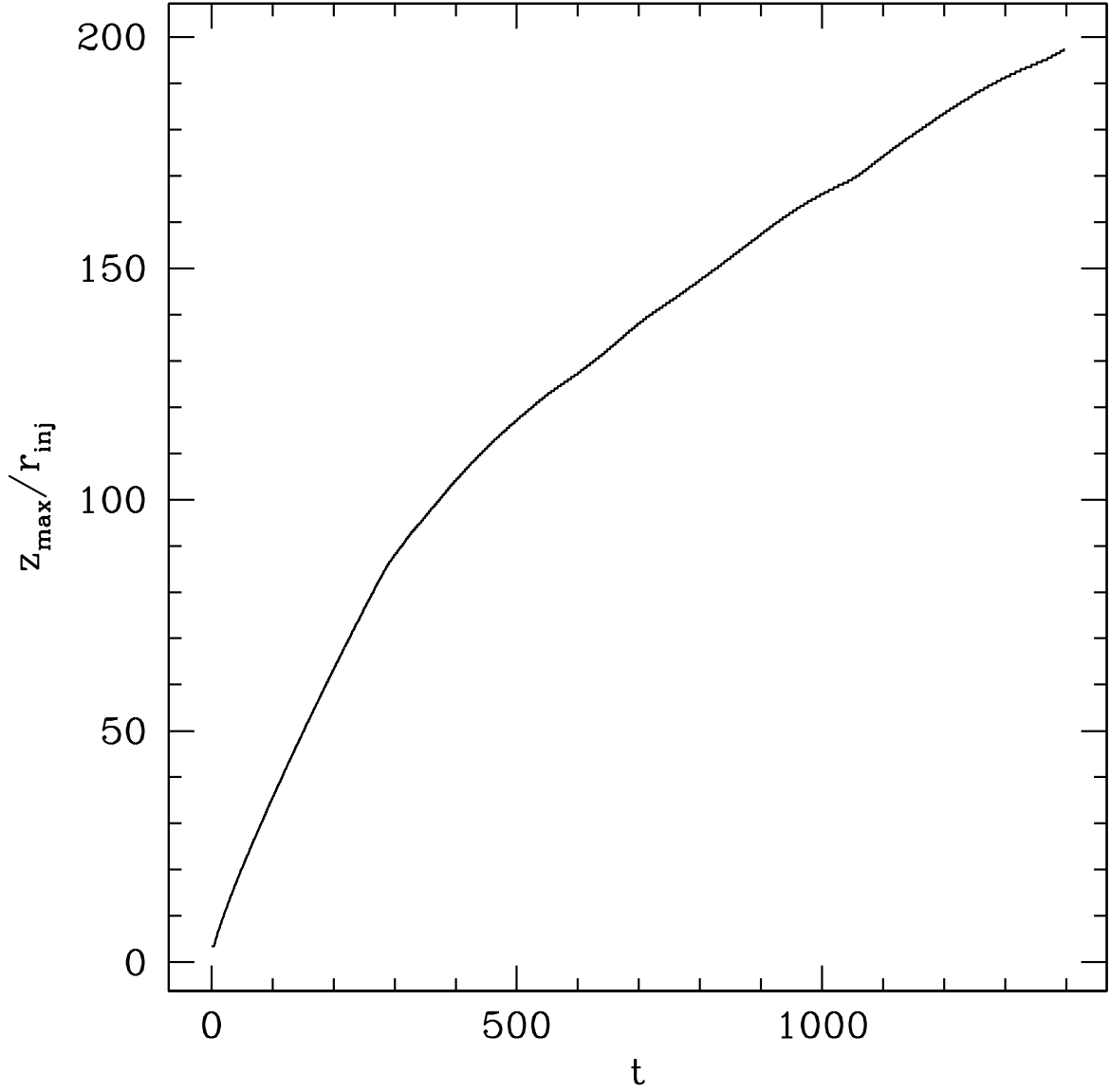}
\caption{The location of jet front along the $z-$ direction as a function of 
time in the fiducial run. The time when the jet slows down ($t \sim 300$) 
is consistent with the appearance of instabilities as shown in Fig. \ref{fig:fidjzcompo}.}
\label{fig:fidjetfront}
\end{figure}

To calculate the jet speed, we can follow the jet front and record its location as a function of time.  
Figure \ref{fig:fidjetfront}  shows how the location of jet front changes over time.  Evidently, the 
jet front starts with an almost constant speed $\sim 0.3c$, and then its propagation 
speed changes at around $t \sim 300$, and gradually slows to $\sim 0.1c$.  
There is no slowing-down at the late time.  Compared to the $j_z$ snapshots sequence 
in Figure \ref{fig:fidjzcompo}, the turning point at the jet propagation occurs at the time 
when the nonlinear modes start to grow significantly. 

\subsubsection{Energy Transition}

\begin{figure}
\plotone{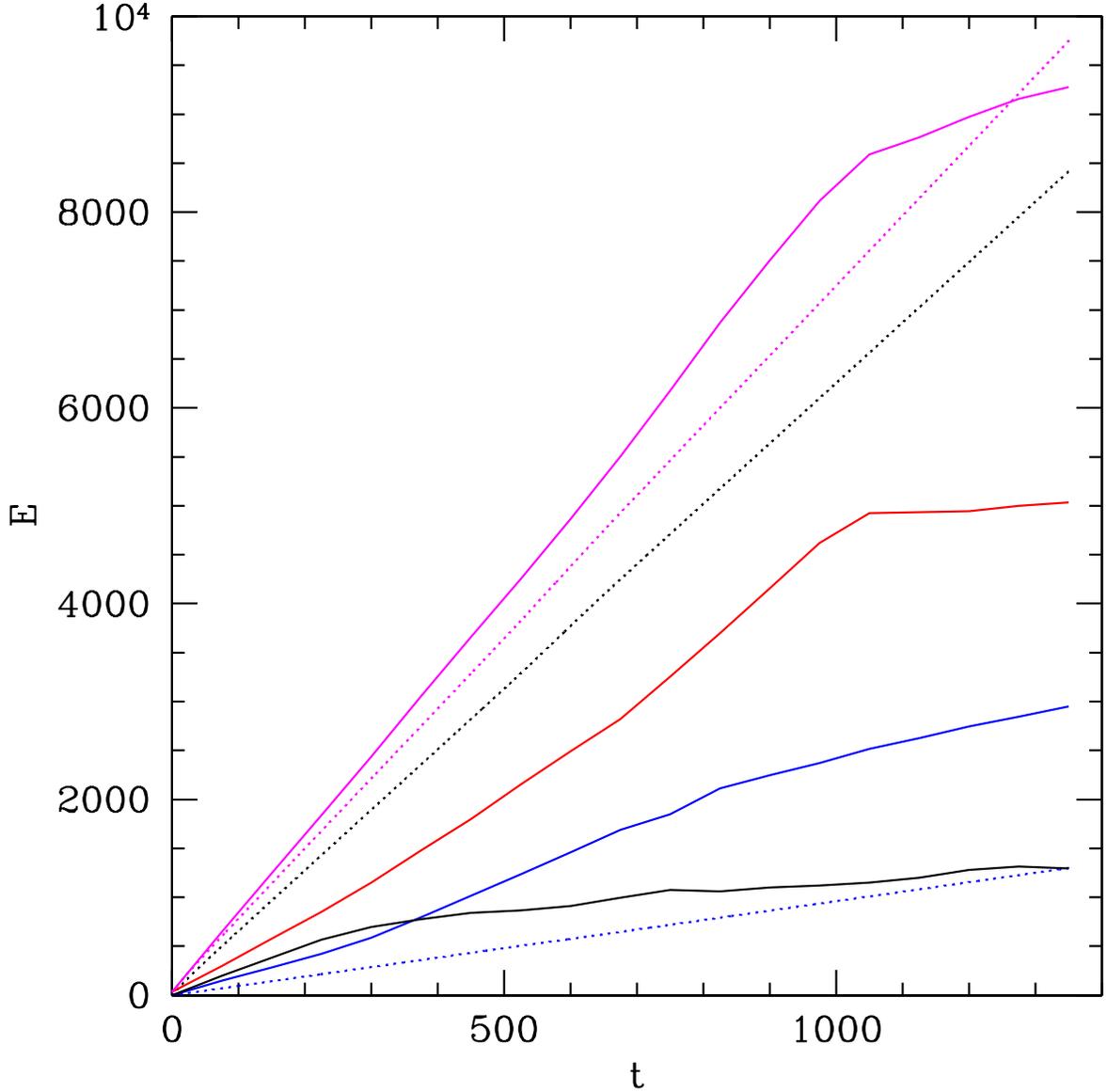}
\caption{Evolution of different energy components of the fiducial run. 
Solid lines denote volume integrated energy and the dotted lines 
denote time and volume integrated injected energy.  
Black solid: $E_{\rm B}$; blue solid:$E_{\rm K}$; red solid: $E_{\rm U}$; 
magenta solid: $E_{\rm tot}$. Black dotted: $E_{\rm B, inj}$; 
blue dotted:$E_{\rm K, inj} + E_{\rm U, inj}$; 
magenta dotted: $E_{\rm tot, inj} + E_{\rm tot,0}$. 
The flattening at $t \sim 1000$ is due to energy flowing 
out of the computational domain.
Even though the injected energy is predominantly magnetic, it gets
converted into kinetic and thermal energies. So, the jet appears
as having a large amount of kinetic and thermal energy. Note that the plotted
quantities are volume integrated. In localized regions such as jet's axis,
magnetic energy can still be comparable to other energy components.}
\label{fig:fidener}
\end{figure}

As the current-driven jet propagates further away from the injection region, 
instabilities grow and non-linear structures develop. These features also affect jet energetics, 
which is a central problem in jet physics. In Figure \ref{fig:fidener} we plot the evolution of 
volume-integrated total magnetic energy $E_{\rm B}$, total kinetic energy 
$E_{\rm K}$, total internal energy $E_{\rm U}$, and total energy $E_{\rm tot} = E_{\rm B} + E_{\rm K} +  E_{\rm U} $.  
Note that magnetic energy density $e_{\rm B}$ includes all terms\footnote{In most of our models, 
the first term dominates by being an order of magnitude larger than the other terms.} 
containing magnetic field,  and it has a form of 
$e_{\rm B} = {\bld B}^2/2 +[|{\bld v}|^2|{\bld B}|^2-({\bld v}\cdot {\bld B})^2]/2 $. 
For the kinetic energy density, we have excluded the rest mass energy, therefore 
$e_{\rm K} = (\gamma -1)\gamma\rho$. The internal energy density is $e_{\rm u} = p/(\Gamma -1) $. 
As a reference, we have also plotted the time and volume integrated injected magnetic 
energy  $E_{\rm B, inj}$,  injected kinetic energy $E_{\rm K, inj}$, and injected internal 
energy $E_{\rm U, inj}$.  Note that for the injected energy the meaningful diagnostic here is 
to calculate the total injection up to a certain time $t$, 
$E_{\rm inj}(t) = \int_0^{t}\int \dot{E}_{\rm inj} dv dt $.  It is obvious that although 
all energetics are increasing with the constant energy injection, after 
$t \sim 300$, $E_{\rm B}$ increases with a much shallower slope 
compared to the growth of $E_{\rm K}$ and $E_{\rm U}$. 
Before $t \sim 300$, the magnetic energy is larger than the kinetic energy 
but after that kinetic energy takes over. 

We have also monitored the total energy conservation during the simulation. 
In Figure \ref{fig:fidener}, the dotted magenta line represents 
$E^{'}_{\rm tot, inj} =  E_{\rm tot, inj} + E_{\rm tot,0}$,  the total injected energy
(magnetic + kinetic + thermal) plus the initial background energy, 
whereas the solid magenta line represents the sum of various energy components in the simulation
domain.  At the beginning they are quite close to each other, but as the simulation 
progresses, the difference between $E_{\rm tot}$ and  $E^{'}_{\rm tot, inj}$ 
continues to increase. The difference between these two total energies, 
however, is always much smaller than the other energy components in the simulation.  
This energy discrepancy is dominated by numerical errors and the origin of 
these errors in MHD simulations is relatively well known.  For our simulations 
we have used both dual-energy formulation (evolving both internal energy and 
total energy equations) and energy fix after the constrained-transport to preserve 
the positivity of the thermal pressure.  Both procedures break the total energy 
conservation in low pressure region and introduce energy error by a small amount.  
In addition, we find that these errors decrease gradually when we increase
the numerical resolution. 
The sudden change in $E_{\rm tot}$ after $t \sim 1000$ is because 
the expansion has reached the computational domain boundaries and materials 
are flowing out of the box.

\begin{sidewaysfigure}
\centering
\scalebox{0.9}{\includegraphics{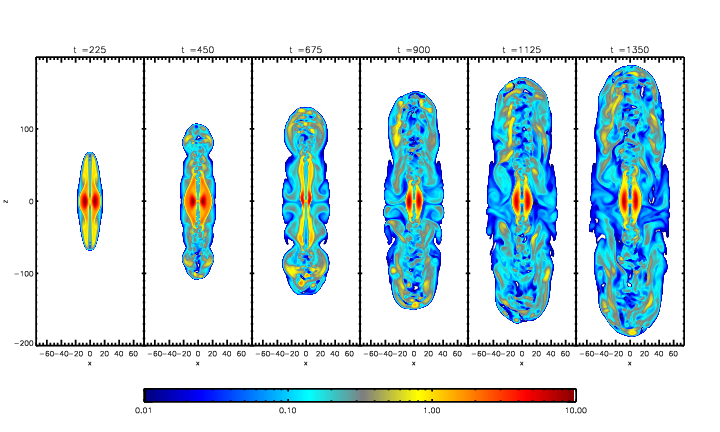}}
\caption{Snapshots of $\sigma$ for the fiducial model. Similar to Fig. \ref{fig:fidjzcompo},
snapshots are taken from $t = 225, 450, 675, 900, 1125, 1350$, at $y = 0$ plane. 
Large $\sigma$ indicates magnetic energy domination. 
}
\label{fig:fidsigmacompo}
\end{sidewaysfigure}

Our numerical model therefore gives an example of transferring jet's magnetic energy 
into kinetic energy as jet propagates.  The magnetization parameter $\sigma$,  
which we have chosen here as the ratio of Poynting energy flux to the kinetic energy 
flux\footnote{Other forms of $\sigma$ exist. Note that the factor of $4\pi$ 
has been absorbed in our numerical representation of the magnetic field.}, 
is $\sigma \equiv F_{\rm Poynting}/F_{\rm P} = B^2/4\pi \gamma ^2 \rho c^2$.  
In Figure $\ref{fig:fidsigmacompo}$ we plot several snapshots of  
$\sigma$ at the $y=0$ plane.  As the jet propagates from its core region, 
the magnetically dominated region has been kept to be a region with a nearly
constant extent $|l_z| \leq 50 ~r_{\rm inj}$.  At late time, as the instability 
causes the jet fields to have more random and small structures, the jet can be 
seen in a more or less kinetically dominated state.  Therefore, our numerical 
model illustrates a jet which contains a near-region with a $\sigma \gg 1$ and 
a far-region with a $\sigma \ll 1$.  The jet does not stop nor get destroyed 
after this transition occurs.  The energy transition is likely 
a result of current-driven instabilities.  

\subsubsection{Current-Driven Instabilities}

\begin{sidewaysfigure}
\centering
\scalebox{0.9}{\includegraphics{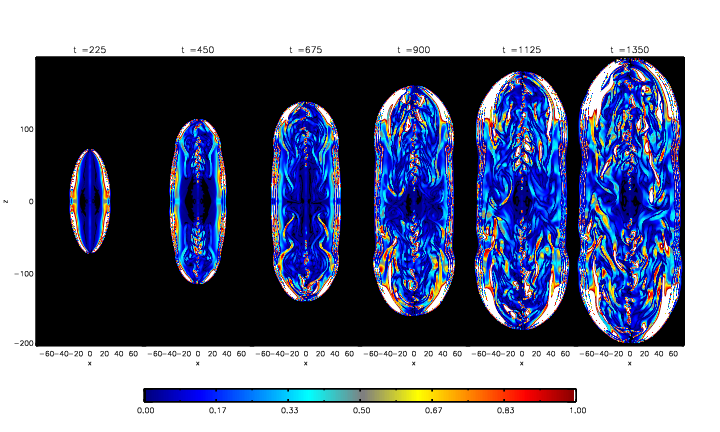}}
\caption{Snapshots of $q$ for the fiducial model. These snapshots are taken from $t = 225, 450, 675, 900, 1125, 1350$, at $y = 0$ plane.  $q<1$ denotes where the current is unstable to the kink mode.
}
\label{fig:fidqcompo}
\end{sidewaysfigure}

In this section we give more details of the CDIs in the fiducial model. 
The primary candidate for CDIs is the kink instability.
According to Kruskal-Shafranov criterion \citep{kado66},  a cylindrical MHD plasma 
with a constant current density $j_z$ in a confined radius is unstable to 
kink modes when $q =  2\pi rB_p/(L_zB_\phi) < q_{\rm crit}$, where $r$ is the 
cylindrical radius, $B_p$ is the poloidal component of the magnetic field 
which is parallel to the axis of the cylinder,  $B_\phi$ is the toroidal field, 
and $L_z$ is the plasma column length. For ideal MHD, $q_{\rm crit} =1$, 
for RMHD, this number is a few \citep{narayan09}.  This instability criterion 
indicates that when the jet is dominated by $B_{\phi}$, the jet will be unstable 
to the $m=1$ kink mode.  This is indeed what we have observed in our simulations. 
In Figure \ref{fig:fidqcompo} we have plotted $q$ at different times in 
the fiducial run, where we have chosen $L_z$ to be the height of the jet at the time. 
We can see that most of the  near-axis and $|l_z| < 50r_{\rm inj}$ 
region with large-current has $q<1$ throughout the simulation. 
Note that the Kruskal-Shafranov criterion is derived from the highly 
ideal situations and we should concentrate on the near-axis region where 
the large current is confined.  The growth of CDIs  
is responsible for the slow-down of the jet front and 
facilitates the energy-transition process.  For the physical parameters 
in our model, this growth period is $\gtrsim 100$ yrs. 

One way to quantify the growth of the nonaxisymmetric modes is to 
calculate the power in the current using Fourier transform
\begin{equation}
f(m,k) =  \frac{\int_{r_{\rm min}}^{r_{\rm max}}\int_{0}^{2\pi}\int_{ z_{\rm min}}^{z_{\rm max}}  |{\bld J}|e^{i(m\phi +kz)} r dr d\phi dz}{\int_{r_{\rm min}}^{r_{\rm max}}\int_{0}^{2\pi}\int_{ z_{\rm min}}^{z_{\rm max}} r dr d\phi dz} 
\end{equation}
where $|{\bld J}|$ is the amplitude of the current density and the integration 
is over a cylindrical volume which encloses the current. In our calculation, 
we have used $r_{\rm min} = 0$, $r_{\rm max} =10r_{\rm inj}$, 
$z_{\rm min} = 0$, and $z_{\rm max} = 200r_{\rm inj}$.   
$m$ is the azimuthal mode number and $k = 2\pi/\lambda$ is the 
vertical wavenumber where $\lambda$ is a characteristic wavelength. 
The volume-averaged mode power in the current amplitude $|J|$ is then
\begin{equation}
P(m,k) = |f(m,k)|^2 = \{{\rm Re}[f(m,k)]\}^2 +\{{\rm Im}[f(m,k)]\}^2, 
\end{equation}
where ${\rm Re}[f(m,k)]$ and ${\rm Im}[f(m,k)]$ are the 
cosine and sine Fourier transformations of $|{\bld J}|$, respectively.

\begin{figure}
\plotone{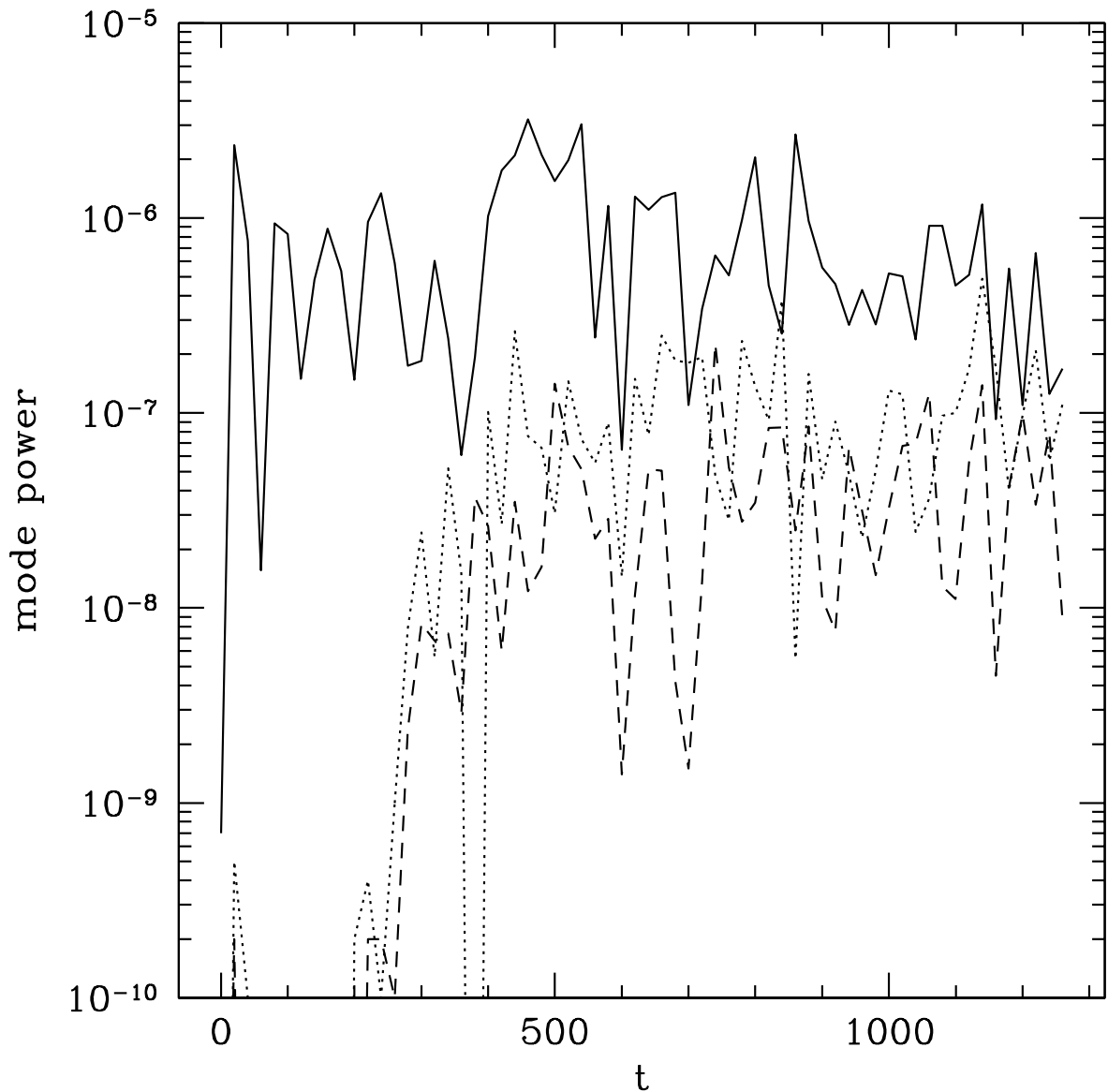}
\caption{Evolution of various mode power in the current distribution
for the fiducial run. Solid lines: $m =0$; dotted lines: $m=1$; dashed lines: $m=2$. 
The axisymmetric component remains dominant throughout the jet evolution.
The non-axisymmetric modes show exponential growth but 
relatively low saturation level at the nonlinear stage.}
\label{fig:fidpow}
\end{figure}

In Figure \ref{fig:fidpow} we plot the time evolution of $P(m,k)$ for the 
$m=0,1,2$ components for the fiducial run. 
For $k$, we have chosen $\lambda = 20 r_{\rm inj}$ for the characteristic 
wavelength (we have examined other wavenumbers and found 
they experience similar exponential growth). 
The $m=0$ component dominates throughout the run, 
although at late times the power in the nonaxisymmetric components 
has grown to be close to the power in the $m=0$ mode. 
The dominant nonaxisymmetric mode is the $m=1$ mode, 
and there is an exponential growth period between 
$t \sim 300 -500$.  After $t\sim 500$, the power in non-axisymmetric 
modes continues to grow, but at a rate which is much slower. 
There is also substantial power in the $m=2$ mode.  Note that 
the background perturbations affect the onset time of significant
growth: we have found that in another simulation with $50\%$ random 
background density perturbations, the onset time has changed 
significantly to about $t \sim 100$.  

We have also observed magnetic Kelvin-Helmholtz instabilities due to 
the large shear that exists at various regions in the jet. 
The characteristic ``cat eye" features can be observed at the jet 
front (e.g. see current near  $z \sim 50 r_{\rm inj}$ in $j_z$ 
slice at $t = 450$ in Figure 1). 

It is noteworthy that although instabilities occur in our models,  
the jet does not get totally disrupted and continues to propagate with an 
almost constant speed.  This is partially due to the  
constant magnetic flux injection which continually drives the jet. 
The fact that the power in $m>0$ modes remaining smaller than the 
power in $m=0$ mode during the nonlinear stage is consistent with 
the non-disruption of the jet. We will discuss the possible explanation for
stabilization in \S 4.

\subsection {Effect of $\alpha$}

The detailed properties of current-driven jets depend on the model parameters, 
one of which is the $\alpha$ parameter that represents the ratio of toroidal to 
poloidal fields. Effects of other parameters on the jet propagation will be examined 
in future studies.

In this simulation we use a higher $\alpha =40$, which gives a stronger
toroidal field injection.  In order to make comparison with the fiducial run, we try to 
keep the same magnetic energy injection rate, we have used a smaller magnetic 
field injection coefficient $B_{\rm inj,0} = 0.054$.  We found the jet propagates 
faster using this injection field configuration. We therefore have used a bigger 
vertical box extent of $L_z = 800$ while keeping  $L_x = L_z = 150$ in order to 
accommodate the jet for the same run duration $t_f = 1500$.  We have also 
increased the grid size to $300\times 300\times 1600$ to keep the same 
resolution as that used in the fiducial run.

\begin{sidewaysfigure}
\centering
\scalebox{0.9}{\includegraphics{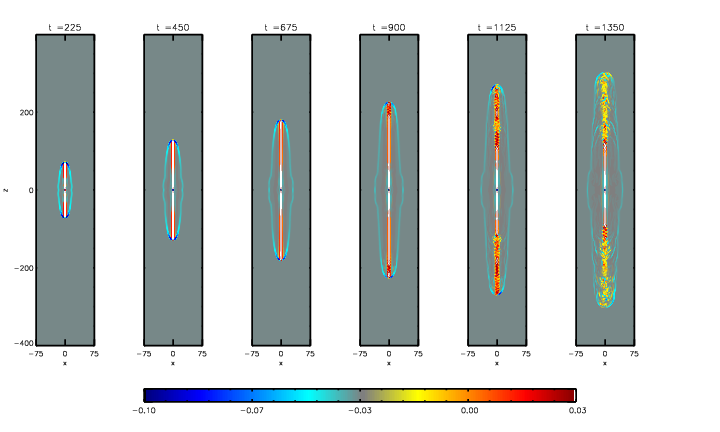}}
\caption{Similar to Fig. \ref{fig:fidjzcompo} but with snapshots of $j_z$ for 
the $\alpha = 40$ model. 
These snapshots are taken from $t = 225, 450, 675, 900, 1125, 1350$, 
at the $y = 0$ plane. The jet is much strongly collimated, presumably
due to the stronger $B_\phi$ injections. 
}
\label{fig:alpha40jzcompo}
\end{sidewaysfigure}

Figure \ref{fig:alpha40jzcompo} plots $y=0$ slices of the z component of current 
density $j_z$ at different times. Notice that the vertical size is twice as that in the 
fiducial run, then this jet definitely moves much faster than the fiducial jet.  
Compared to the $\alpha = 10$ run, the non-linear features appear at a 
much later time, at a higher $z$ location, takes longer to grow, and the jet 
also has a leaner shape. In the $\alpha = 10$ run, the non-axisymmetric 
modes appear to grow exponentially from $t\sim 300-500$, while here the 
instabilities do not start significant growth after $t \sim 500$. The current 
is also more concentrated toward the z-axis, most likely due to increased 
hoop pressure resulted from the larger $B_\phi$ component.

\begin{figure}
\centering
\begin{tabular}{cc}

\epsfig{file=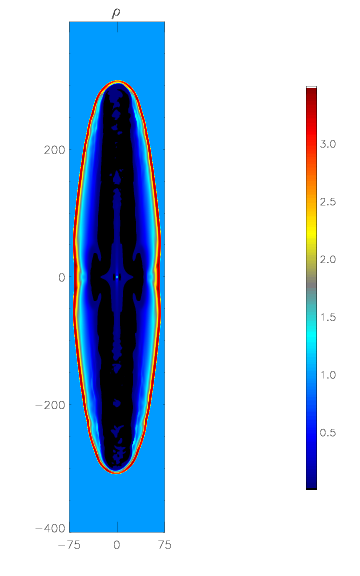, width=0.4\linewidth, clip=} &
\epsfig{file=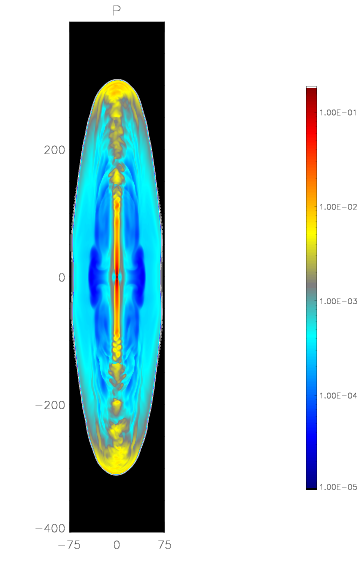, width=0.4\linewidth, clip=} \\
\epsfig{file=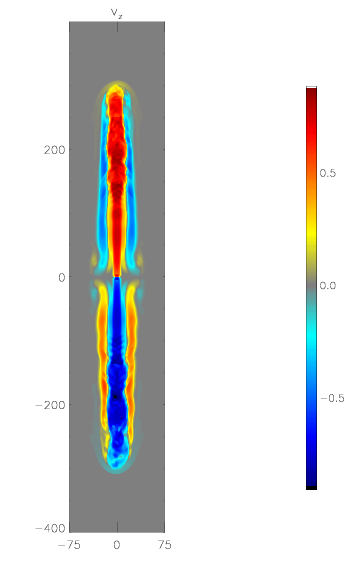, width=0.4\linewidth, clip=} &
\epsfig{file=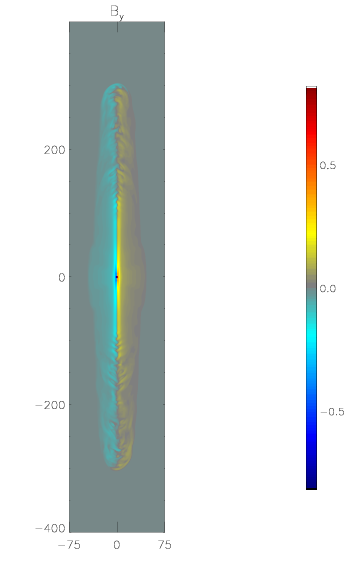, width=0.4\linewidth, clip=} \\

\end{tabular}
\caption{Similar to Fig. \ref{fig:fidvars} but with 
snapshots of $\rho, P, v_z, B_y$ at late time for the $\alpha =40$ run. 
These snapshots are taken from $t = 1350$, at $y = 0$ plane.  
}
\label{fig:alpha40vars}
\end{figure}

Figure \ref{fig:alpha40vars} shows snapshots of $y=0$ plane cut-through 
for $\rho, P, v_z, B_y$ at late time $t = 1350$.  Despite the more 
elongated jet shape,  all the plotted quantities show qualitatively similar 
behaviors compared to the smaller $\alpha$ run. The Lorentz factor 
continues to increase over time and the highest Lorentz factor achieved 
in this run is about $\gamma \sim 2.4$. We suspect this number will 
increase more as the jet has not developed much non-linear features 
at the end of run.  However, it is not clear what determines the terminal 
$\gamma$ in our models, as it needs a much bigger computational 
domain size as well as longer simulation run time.  

\begin{figure}
\plotone{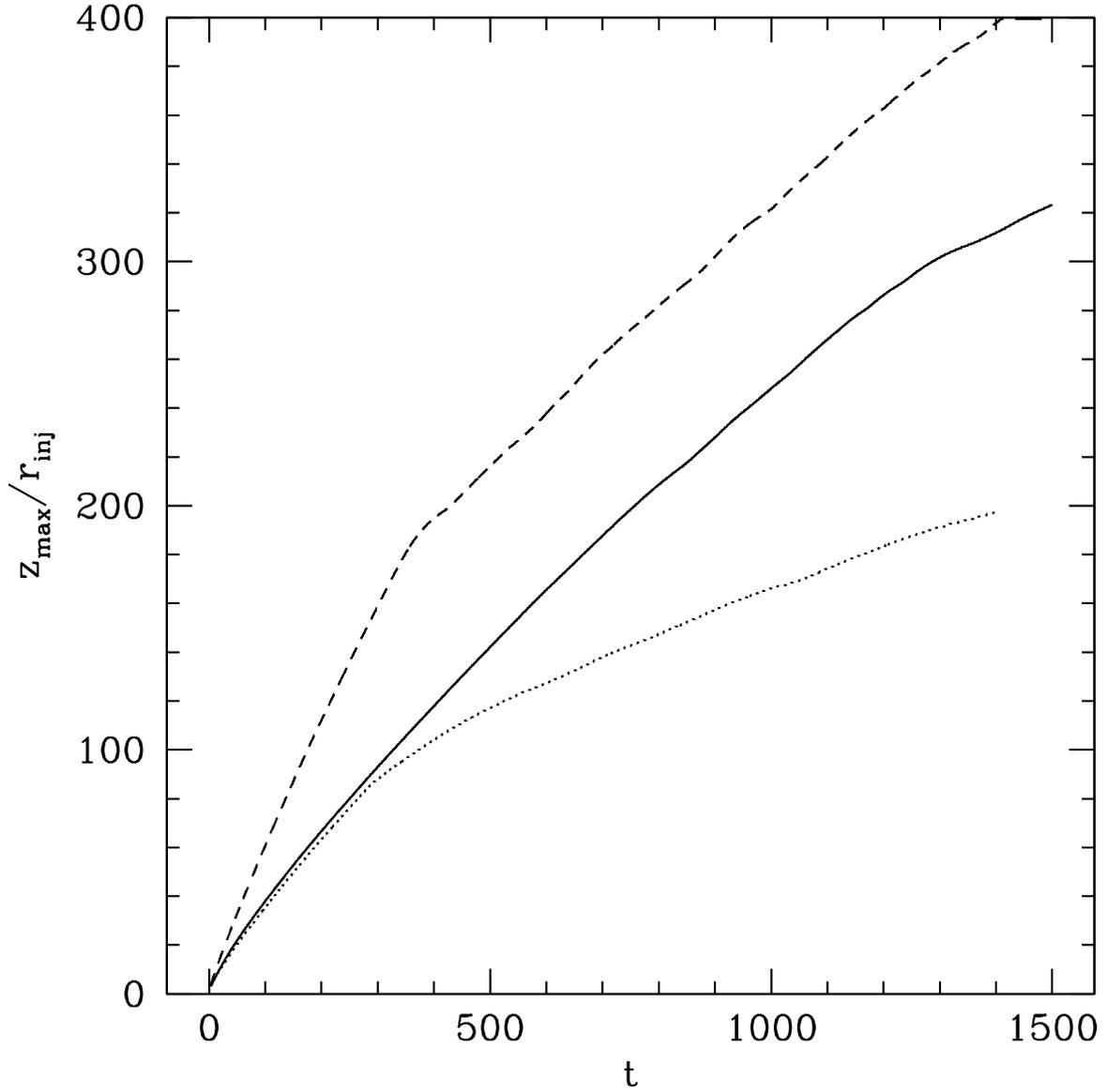}
\caption{The location of jet front as a function of time in the $\alpha=40$ runs. 
Jet slows down after the non-axisymmetric modes become significant 
compared to the axisymmetric mode. Solid: $\alpha = 40$ with the same 
total magnetic energy injection rate as the fiducial run; dotted: $\alpha =10$ 
fiducial run; dashed: $\alpha =40$ but with a larger magnetic energy 
injection rate. }
\label{fig:alpha40jetfront}
\end{figure}

Figure \ref{fig:alpha40jetfront} illustrates the propagation of jet front for 
$\alpha = 40$ case.  The slowing down of jet front does not occur until 
$t \sim 1200-1300$, much later compared to the smaller alpha case.  
Although the injected magnetic energy rate is the same, the jet propagates 
with a larger bulk velocity because the dominant toroidal components, 
consistent with predictions by the magnetic tower models
(see discussion in the \S 4).

\begin{figure}
\plotone{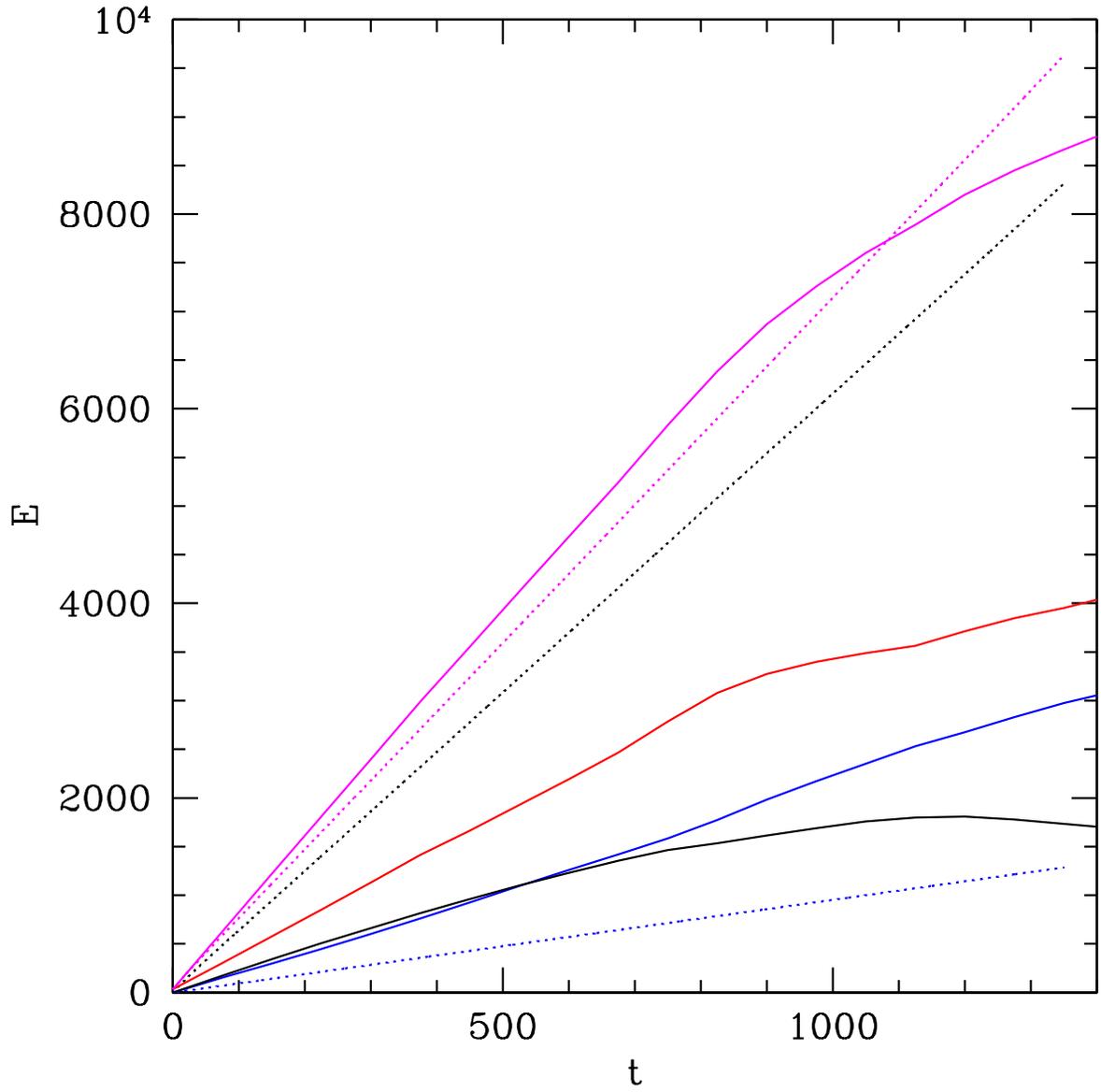}
\caption{Energetics of the $\alpha = 40$ run. Color scheme is the same as in Figure \ref{fig:fidener}. }
\label{fig:alpha40ener}
\end{figure}

\begin{sidewaysfigure}
\centering
\scalebox{0.9}{\includegraphics{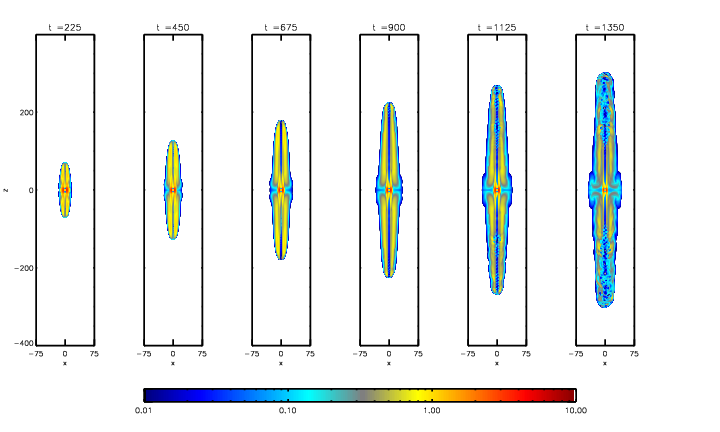}}
\caption{Snapshots of $\sigma$ for the $\alpha = 40$ model. 
These snapshots are taken from $t = 225, 450, 675, 900, 1125, 1350$, at $y = 0$ plane. 
}
\label{fig:alpha40sigmacompo}
\end{sidewaysfigure}

We have observed similar behavior for total energetics in this model as 
in the $\alpha = 10$ case, as shown in Figure \ref{fig:alpha40ener}. 
Similar to the fiducial run, the total kinetic energy takes over the magnetic 
energy after the instabilities grow, and both the kinetic energy and internal 
energy increase with the continuous conversion of magnetic energy 
into these two energies. $E_{\rm K} > E_{\rm B}$ occurs at a later time 
compared to the fiducial run, consistent with the onset of non-linear features. 
At the end of the simulation, the total $E_{\rm K}$ is quite similar to 
the $E_{\rm K}$ in the fiducial run,  $E_{\rm B}$ is $\sim 34\%$ larger than 
that in the fiducial run, and the total internal energy is $\sim 27\%$ smaller 
than in the fiducial run.  This smaller dissipation is also consistent with the 
later onset of non-linear features.  The smaller energy transition can 
also be seen from the magnetization parameter $\sigma$ images. 
Figure \ref{fig:alpha40sigmacompo} shows $\sigma$ at $y=0$ slices at 
different times for this run.  It is clear that, when compared to the fiducial run, 
the energy transition occurs mainly at a later time too, consistent with the 
onset time for the significant non-linear interactions. This means for the same 
amount of total magnetic energy injection, when $\alpha$ is larger, the energy 
transition will occur further away from the jet launching location.  

\begin{sidewaysfigure}
\centering
\scalebox{0.9}{\includegraphics{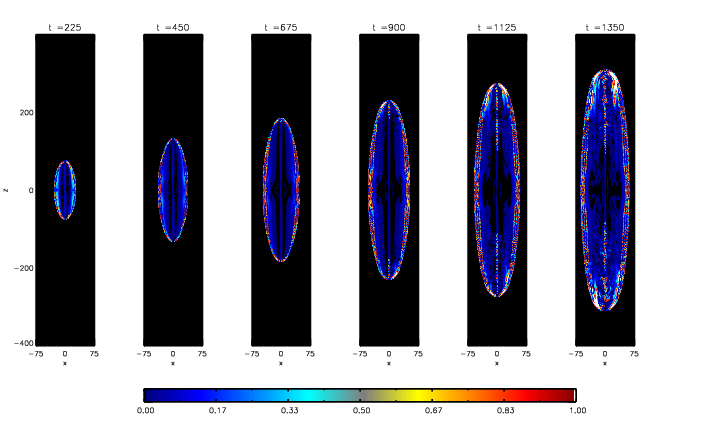}}
\caption{Snapshots of $q$ for the $\alpha = 40$ model. 
These snapshots are taken from $t = 225, 450, 675, 900, 1125, 1350$, at $y = 0$ plane.  $q<1$ denotes when the current is unstable to the kink mode.
}
\label{fig:alpha40qcompo}
\end{sidewaysfigure}

How about CDIs? Figure \ref{fig:alpha40qcompo} plots the snapshots of 
value of $q$ for the kink instability limits at $y=0$ slices. For a certain 
cylindrical current, when $\alpha$ increases, the $q$ value decreases 
for the same cylindrical shape. Therefore, the jet will still be unstable 
due to the kink instabilities, and this is what we have observed here. 

\begin{figure}
\plotone{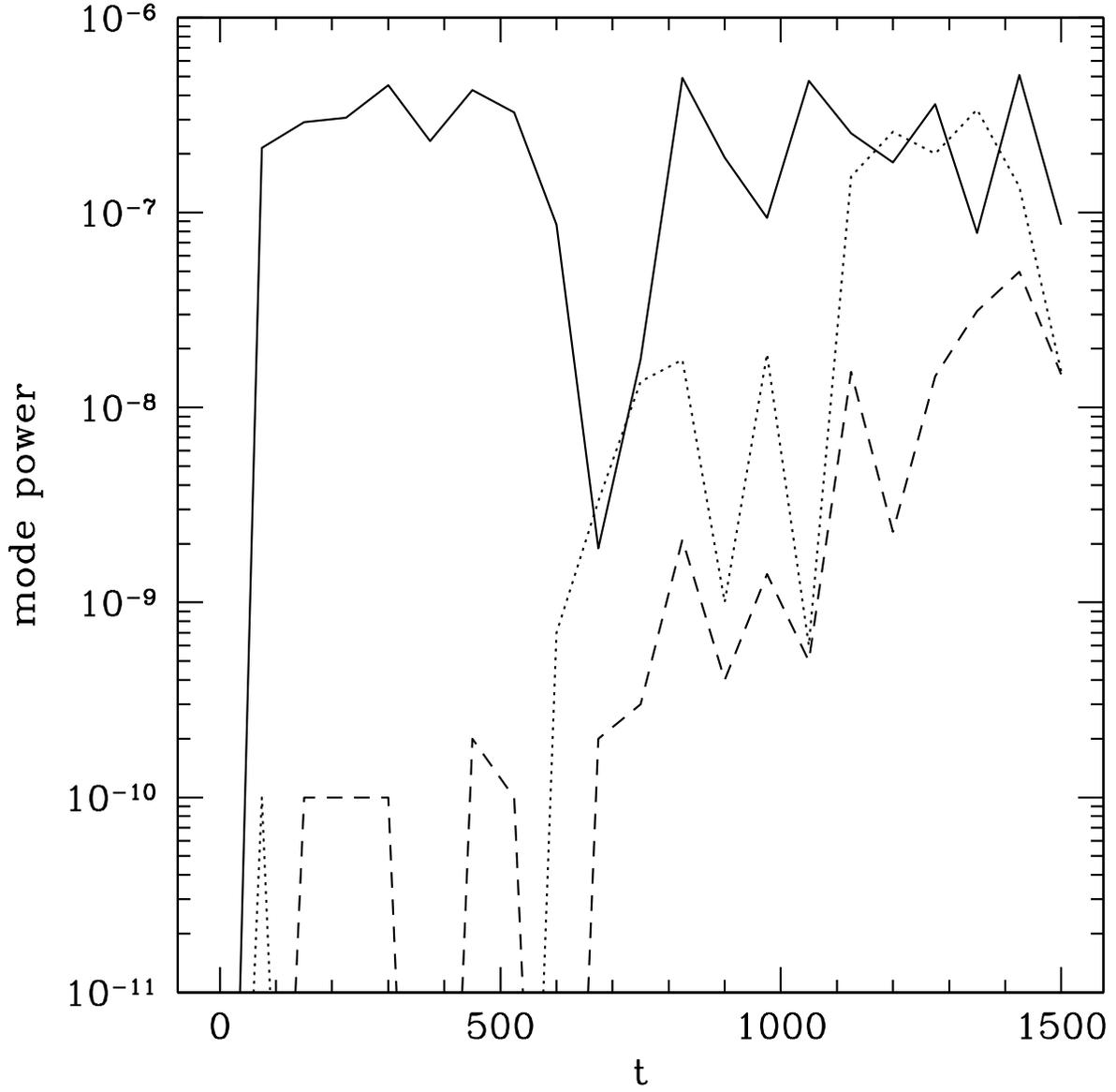}
\caption{Evolution of mode power in the current for the $\alpha = 40$ run. Solid lines: $m =0$; dotted lines: $m=1$; dashed lines: $m=2$. }
\label{fig:alpha40pow}
\end{figure}

To see the detailed interplay between axisymmetric and non-axisymmetric 
modes, we have calculated the power of first few modes in this model. 
Figure \ref{fig:alpha40pow} shows the growth of mode power of the amplitude 
of current for this run. Similar to the lower $\alpha$ model, the dominant 
non-axisymmetric mode is the $m = 1$ kink mode. Throughout the simulation 
the axisymmetric $m=0$ mode dominates, although the $m = 1$ mode 
almost grows to a similar magnitude at the late time, which introduces the 
non-linear behaviors. However, the growth rates of non-axisymmetric modes 
are smaller compared to the smaller $\alpha$ case. 
This is somewhat surprising as the larger $\alpha$ is expected to lead to a 
stronger instability.  
One possible explanation is that, while the linear analysis for the growth rate 
of kink instability is based on the ideal setup of a constant cylindrical current 
with well-defined geometry and fixed boundaries,  here we are dealing with 
an evolving jet with continuous magnetic injection at the center and the jet itself 
is fast propagating in the vertical direction and expanding in the transverse directions. 
Therefore instability analysis from ideal plasma physics derivation may not be 
applied directly to our evolving system.  Further discussions on this result 
are given in \S 4. 

To understand the dependence of the CDI's on-set on injection parameters, 
we also make a run where the poloidal field injection rate is the same as the 
fiducial run ($B_{\rm inj, 0} = 0.2$) while keeping $\alpha =40$ (hence
a higher total magnetic energy injection rate), we find that instabilities grow at 
a rate that is more close to that in the fiducial run, and the jet front propagation 
speed turn-over occurs earlier, at $t \sim 400$ (see the dashed line in 
Figure \ref{fig:alpha40jetfront}). This indicates that the growth of CDIs and the onset 
of nonlinear features in these propagating current jet systems are a complex process 
probably depending more on the parameters for the magnetic field injection profile
(both magnitude and shape), and we will explore this more in the future. 

\subsection{Effect of a Disk} 

Our simulations show that the magnetic structure expands both along the
$z-$axis and sideways. As the jet is a consequence of accretion, and in the
spatial scales we are considering, the accretion disk should surround and extend
into the injection region. In this section, we use a toy model to investigate the 
effect of possible disk confinement and whether the instabilities will still occur 
when there is a gas-pressure-supported disk at the jet base. All the jet parameters 
are the same as in the fiducial run.

The reason to choose a gas pressure-supported disk instead of a rotation-supported 
disk is mainly of numerical consideration.  For a more physical accretion disk with 
rotation, the simulation requires a much smaller time step, in order to resolve the 
disk rotation. We therefore choose a gas pressure supported disk which is initially 
in a hydrostatic equilibrium, and this is numerically much easier than evolving 
a rotating disk. We are not modeling the accretion process itself, but focusing 
on how the gas pressure will confine the jet shape and whether the disk will 
affect the instabilities.

We have solved the effective gravitational potential $\Phi_{\rm eff}$ which is 
able to hold a gas disk with a density distribution 
\begin{equation}
\rho (r, z) = \rho_{\rm bkg} + \frac{\rho_0}{(1+r/r_0)^{3/2}}\exp{(-\frac{z^2}{2H^2})},
\end{equation}
where the disk is centered at $x=y=z=0$, $r = (x^2 + y^2)^{1/2}$,
$\rho_0$ is the characteristic disk midplane (defined as $z=0$) 
gas density, $r_0$ is a characteristic disk radius, and $H$ is the disk scale height. 
When choosing $\rho_0 \gg \rho_{\rm bkg}$, the first term in the density equation can be omitted. $\Phi_{\rm eff}(r,z)$ can be solved by considering the Euler equation in the radial and vertical directions. Because there is no rotation and we seek steady-state solutions, the equations are a set of partial differential equations (PDE) of a simple form:
\begin{equation} 
\begin{matrix}
\del_{r} \Phi_{\rm eff}(r,z) = -\frac{1}{\rho (r,z)}\del_{r}p, \\
\del_{z} \Phi_{\rm eff}(r,z) = -\frac{1}{\rho (r,z)}\del_{z}p.
\end{matrix}
\end{equation}

Assuming a simple, constant sound speed $c_{s0}$, the solution of the above PDE can be obtained by integrating separately along $r$ and $z$ directions. $\Phi_{\rm eff}(r,z)$ has a form

\begin{equation} 
\Phi_{\rm eff}(r,z) = c_{s0}^2[\ln(1+(\frac{r}{r_0})^{3/2}) + \frac{z^2}{2H^2}].
\end{equation}
For simplicity we have omitted the constant term. Including a non-trivial $\rho_{\rm bkg}$ 
term in the disk density distribution makes solving $\Phi_{\rm eff}(r,z)$ much more complex.  

To set up this disk, we have chosen $\rho _0 = 100 $ which is much greater than 
the background density in the whole simulation box.  We choose $r_0 = 10 r_{\rm inj}$, 
$H = r_{\rm inj}$, and the same sound speed used for the background gas. 
The inner edge of the disk is set at $r_{\rm inj}$ and outer edge of the disk 
extends to the edge of the box. The disk is thin in most of the regions 
except in the inner few $r_{\rm inj}$.  We have tested our effective 
gravitational potential $\Phi_{\rm eff}(r,z)$ and the associated 
disk density distribution $\rho (r, z)$. In the case of zero injection, 
our disk can indeed be held in a hydrostatic equilibrium by the effective potential. 
After injecting the strong magnetic flux into the center region, the disk cannot 
be retained in its original equilibrium, and will be pushed outward by the 
strong magnetic pressure.  Again, our emphasis of this toy model is to 
test whether the inclusion of a gaseous disk will change the properties of 
the propagating jet, especially the path of the return current profile. 
 
\begin{sidewaysfigure}
\centering
\scalebox{0.9}{\includegraphics{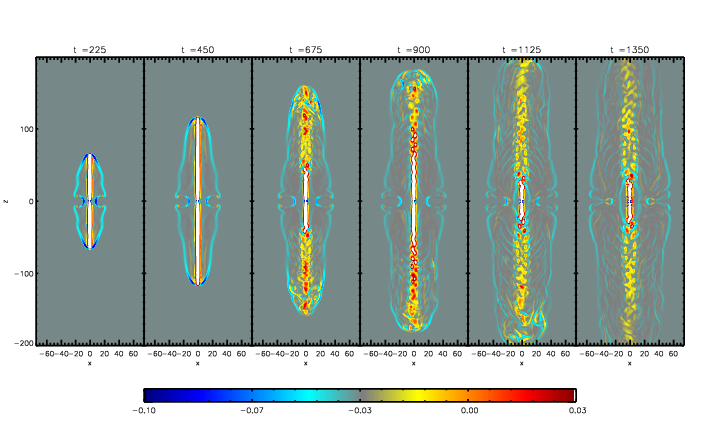}}
\caption{Snapshots of $j_z$ for the disk model. 
These snapshots are taken from $t = 225, 450, 675, 900, 1125, 1350$, at $y = 0$ plane. 
}
\label{fig:dh1jzcompo}
\end{sidewaysfigure}

\begin{figure}
\centering
\begin{tabular}{cc}
\epsfig{file=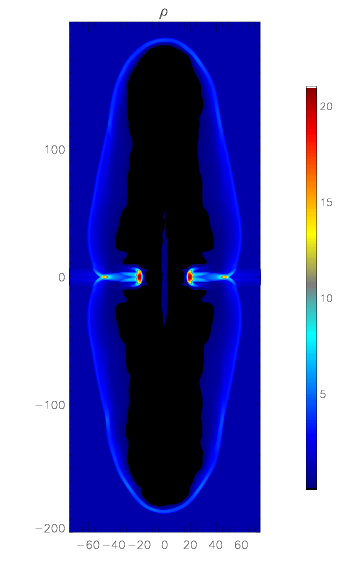, width=0.4\linewidth, clip=} &
\epsfig{file=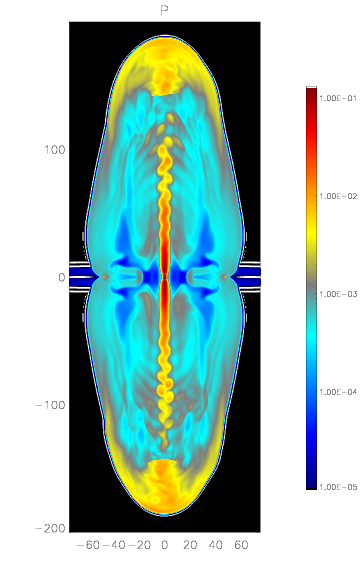, width=0.4\linewidth, clip=} \\
\epsfig{file=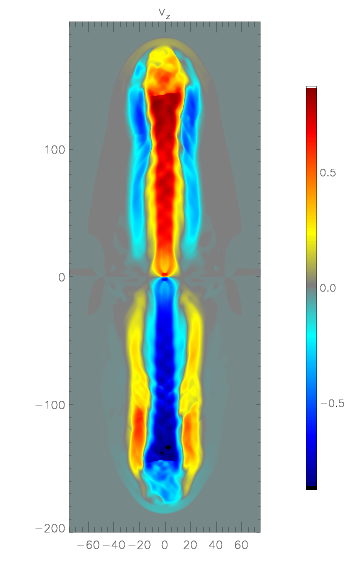, width=0.4\linewidth, clip=} &
\epsfig{file=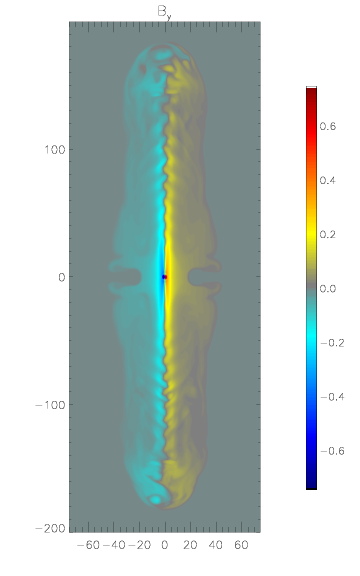, width=0.4\linewidth, clip=} \\
\end{tabular}
\caption{Snapshots of $\rho, P, v_z, B_y$ at late time for the disk run. 
These snapshots are taken from $t = 900$, at $y = 0$ plane.  
}
\label{fig:dh1vars}
\end{figure}

Figure \ref{fig:dh1jzcompo} shows the current density slices at different times when 
including this gas disk. Compared to the fiducial run, near the base of the 
jet ($z \leq 10 r_{\rm inj}$), the jet expands less in the equatorial plane. 
The return current is also much closer to the axis in this region 
(which changes the magnetic field shape more paraboloidal). 
The overall shape of the jet resembles more of an observed astrophysical 
jet in this situation, with an opening angle at its base due to the disk confinement. 
Other quantities are shown in Figure \ref{fig:dh1vars}, which gives 
snapshots of $\rho, P, v_z, B_y$ at the late time. The disk component 
can be clearly seen in these snapshots. The magnetic pressure is 
gradually pushing the disk outward due to the constant flux injection, 
even at the late stage of the simulation: our disk never reaches a static state 
in this model and this is due to the fact that we are not simulating a 
real accretion event here. However, our simple toy model provides a 
glimpse into what a more realistic disk-jet simulation would illustrate in the future.  

\begin{figure}
\plotone{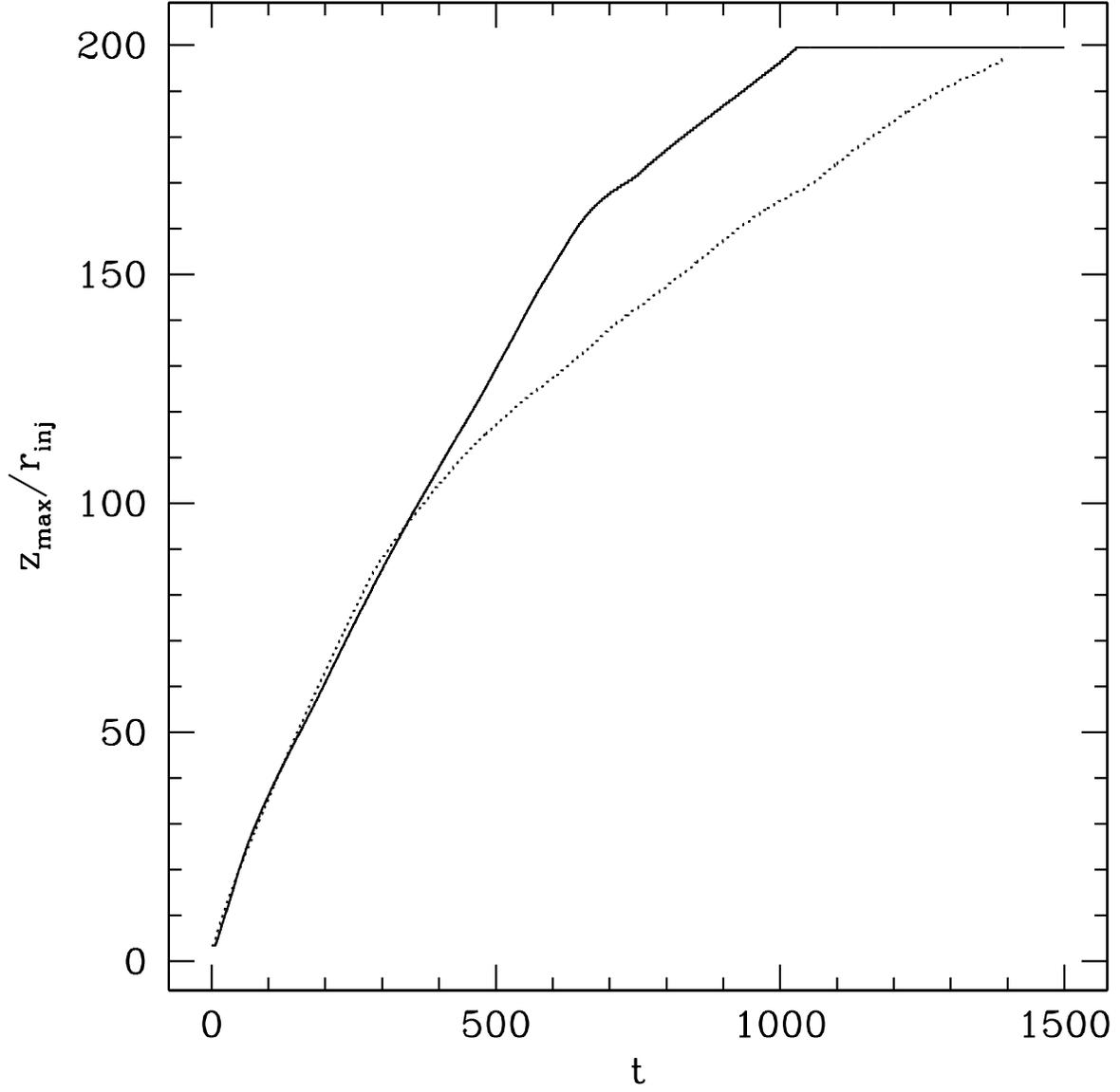}
\caption{The location of jet front as a function of time in the 
disk run (solid line). Jet slows down after the nonlinear 
modes start to grow. Dash line: jet front locations in the fiducial run.}
\label{fig:dh1jetfront}
\end{figure}

More importantly, on the larger vertical distance, the jet displays a very 
similar morphology as in the fiducial run. The jet is well collimated, 
the CDI grows and non-linear features have developed as jet 
propagates beyond a few tens of $r_{\rm inj}$. 
In Figure \ref{fig:dh1jetfront} we have plotted the propagation 
of jet front. It is obvious that jet front has already reached the vertical 
edge of the box at $t = 1000$. The jet front 
propagates with a high speed for a longer duration ($t\sim 450$)
than in the fiducial case. After this stage the jet front propagation slows down but is
still slightly faster than in the fiducial case, most likely due to the extra "pinch'' 
effect at its base.

For instabilities, from instability criterion and mode power analysis 
we find their general properties are quite similar to the fiducial run, 
although the instability growth rate is slightly larger. This is not surprising 
because the instabilities are driven by the injected current, and how they grow 
is a reflection of the intrinsic property of the jet current at large distance, 
rather that the environment confinement provided at its base.

\begin{figure}
\plotone{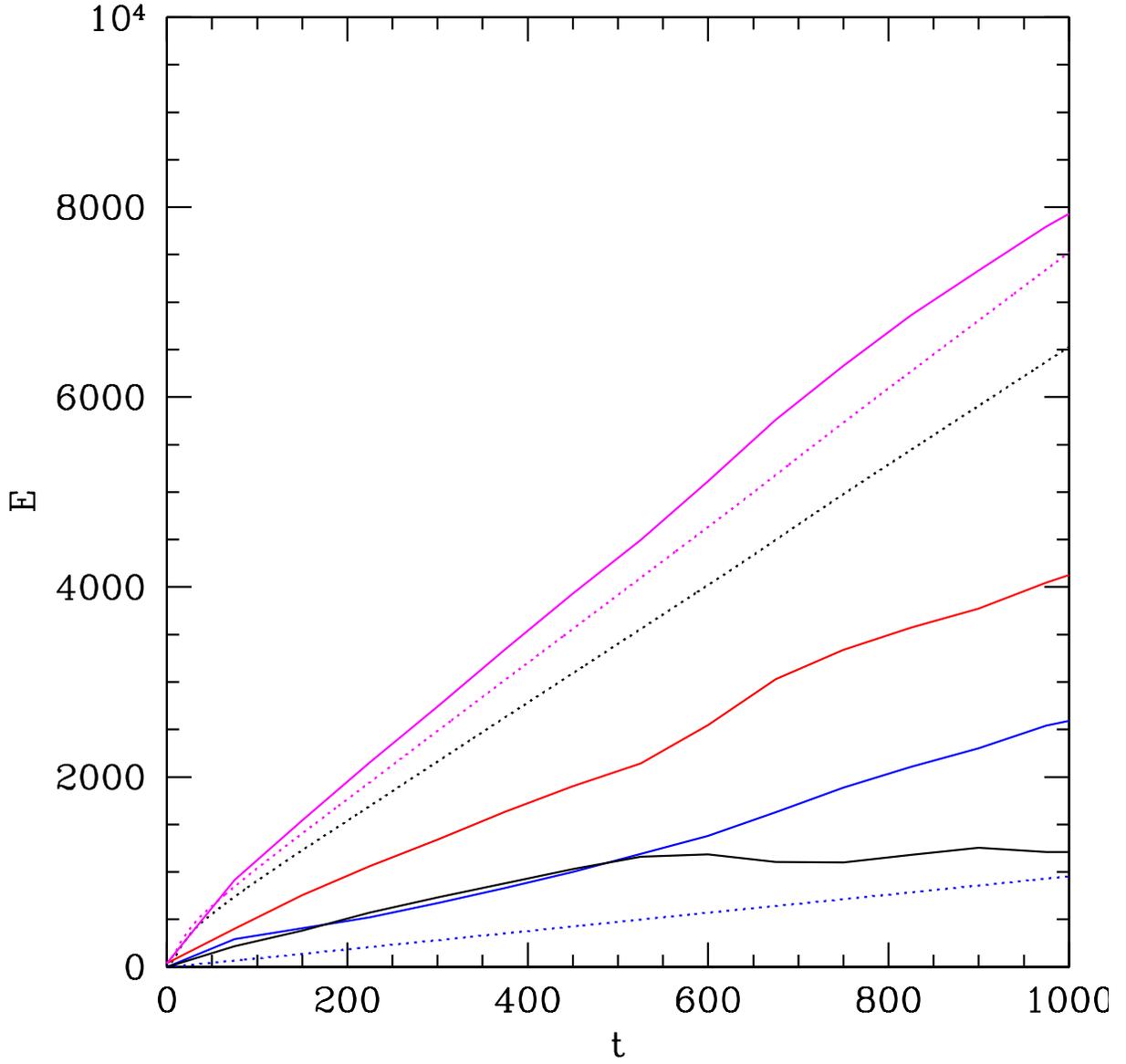}
\caption{Energetics of the disk run.  Color scheme is the same as in Figure \ref{fig:fidener}. }
\label{fig:dh1ener}
\end{figure}

\begin{sidewaysfigure}
\centering
\scalebox{0.9}{\includegraphics{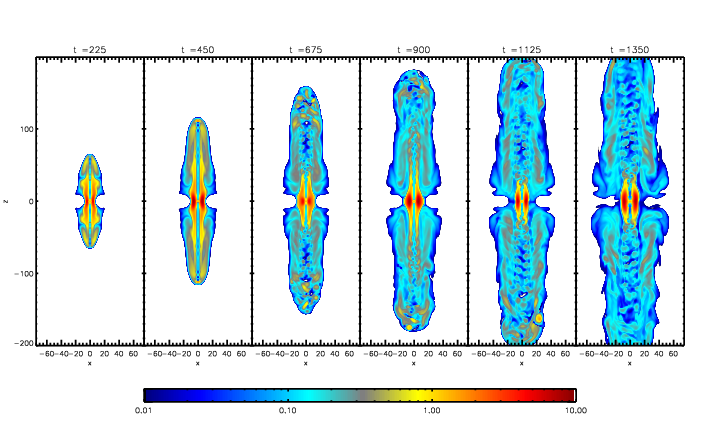}}
\caption{Snapshots of $\sigma$ for the disk model. 
These are taken from $t = 225, 450, 675, 900, 1125, 1350$, at $y = 0$ plane. 
}
\label{fig:dh1sigmacompo}
\end{sidewaysfigure}

For energy transition, Figure \ref{fig:dh1sigmacompo} shows $\sigma$ 
at different time.  This illustrates that, even with a disk, at distances far from the disk and 
injection region, the instabilities introduce large dissipation and magnetic energy 
is transformed into kinetic and thermal energies. We have also calculated the 
evolution of energetics of the total box, as shown in Figure \ref{fig:dh1ener}. 
We get quite similar results compared to the fiducial run: total kinetic energy 
takes over the magnetic energy after the instabilities grow, and both the total kinetic 
energy and the total internal energy increase as magnetic energy is converted 
into these two energies over time.  

We have made additional runs by changing the disk scale height $H$ to a different value 
($H = 5r_{\rm inj}$ which sets up a thicker disk), similar results were obtained.

\subsection{Resolution Study}

\begin{sidewaysfigure}
\centering
\scalebox{0.9}{\includegraphics{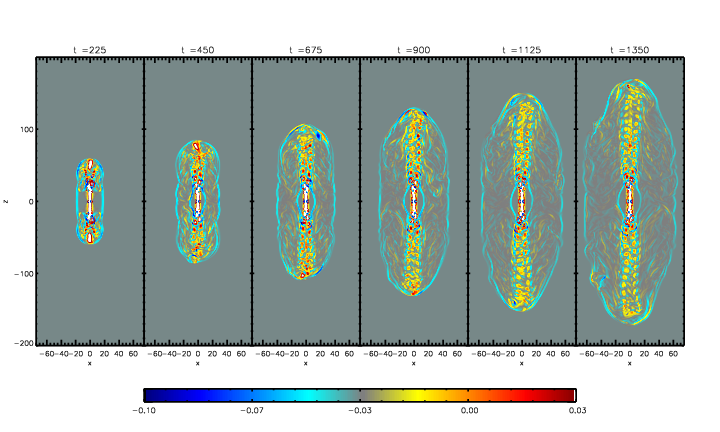}}
\caption{Snapshots of $j_z$ for the fiducial model with higher resolutions. 
Snapshots are taken from $t = 225, 450, 675, 900, 1125, 1350$ at $y = 0$ plane. 
}
\label{fig:hiresjzcompo}
\end{sidewaysfigure}

In order to illustrate the effects of resolution, we have re-run the fiducial case with 
a higher resolution $N_x =  N_y = 450, N_z = 1200$, while keeping all other 
parameters unchanged. 
Figure \ref{fig:hiresjzcompo} shows the $j_z$ current density slices 
at the $y = 0$ plane.  Compared to the fiducial run, the non-linear 
features appear earlier, already apparent at $t \sim 200$. At the late time, 
the jet has a more pronounced  ``spine'', where large scale wiggles in this spine 
are visible near both sides of the jet front. The return current also exhibits 
asymmetric morphology, and extends slightly further away from the axis 
in the equatorial plane. Recently, \cite{mignone10} have studied resolution 
effects in RMHD simulations of jets. They also observed that as jet propagates 
further its trajectory becomes more curved, moving from the central axis.  
This effect is more pronounced in their higher resolution runs.  We note our 
findings are consistent with their results.

Comparing the jet front location at different times for both runs, we find that the 
higher resolution jet propagates first with a similar speed compare to that in 
the fiducial run. Its slowing-down point, however, occurs earlier at $t \sim 150$ 
due to the early onset of the non-linear stage. After $t \sim 150$, the jet propagates 
again with the similar speed as in the fiducial run. This explains why the jet front 
reaches a lower $z$ height compared to the fiducial run at the late time.

Although the resolution does not affect much of the overall jet dynamics, 
it certainly affects the instabilities. From the mode analysis we find that the 
higher resolution simulation also gives an almost doubled growth 
rate for non-axisymetric modes, which causes the current profile to become
nonlinear at $ t \leq 200$.

\begin{sidewaysfigure}
\centering
\scalebox{0.9}{\includegraphics{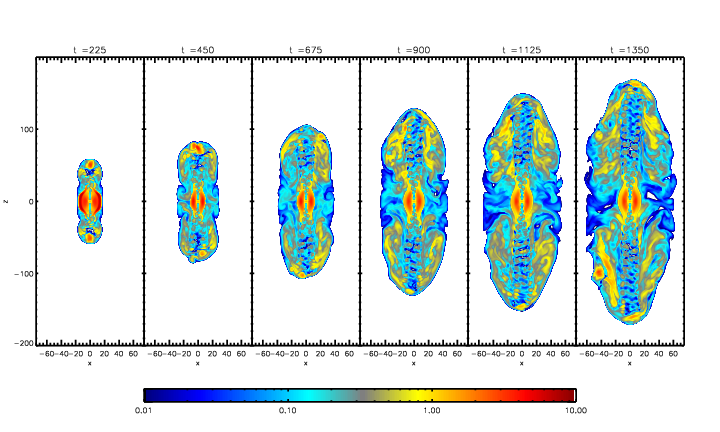}}
\caption{Snapshots of magnetization parameter $\sigma$ for the 
fiducial model with higher resolutions. These snapshots are taken 
from $t = 225, 450, 675, 900, 1125, 1350$, at $y = 0$ plane. 
}
\label{fig:hiressigmacompo}
\end{sidewaysfigure}

Also similar to \cite{mignone10},  we find more and stronger shocks in the 
high resolution run. This introduces more dissipation and gives a larger 
total thermal energy. As a result, we also notice that both the total 
kinetic energy $E_{\rm K}$ and the total magnetic energy $E_{\rm B}$ are smaller 
in the higher resolution run: for example, $E_{\rm B}$ is $\sim 11\%$ smaller 
than that in the fiducial run and $E_{\rm K}$ is $\sim 7.6\%$ 
smaller at $t \sim 600$ when both models are at the non-linear stage. 
The magnetic-to-kinetic energy transition still occurs in the higher resolution run.  
We have plotted $\sigma$ parameter at $y=0$ plane at different times for 
this model, as shown in Figure \ref{fig:hiressigmacompo}. We can see that 
at the ``spine'' region of the jet $\sigma$ is smaller, indicating higher resolution 
leading to a more efficient energy transition. 
 
Lastly, we want to stress that although our higher resolution simulation has 
displayed quantitatively similar behaviors as those in the fiducial run, 
such as the development of CDIs and the energy transition,  our numerical model 
of RMHD jet has not shown signs of convergence. The convergence issue is 
therefore out of the scope of this paper, and needs further investigation.

\section{Summary and Discussion} 

We have carried out new RMHD simulations for Poynting-flux driven jets 
in AGN systems. The computational domain is relatively large so that both the
injected magnetic fluxes and their subsequent evolution are contained well within
the simulation domain. The 
fluxes which are responsible for driving the jet are injected at the center of the 
box, with an injection region size $r_{\rm inj}$. The flux injection rate is continuous
and is taken to be constant. Our injected magnetic fields have an axisymmetric geometry with
close field lines, consisting of a poloidal field plus a dominant toroidal field component. 
We follow the propagation of the jet to a few hundreds of $r_{\rm inj}$ in three dimensions.
We proposed to scale the injection region $r_{\rm inj}$ to 
$\sim 10^3$ gravitational radii of a black hole, thus our simulations could be relevant
to observations of AGN jets on from sub-pc to tens of pc scales. 

We find these jets are well-collimated. They have a concentrated ``spine'' that is roughly 
of the same size of the injection region inside which the majority of the out-going current 
is flowing, along with a significant fraction of the injected poloidal flux. Driven by the
strong magnetic pressure gradient in the $z-$direction, it eventually develops relativistic 
speeds. The magnetic structure also expands transversely, though at a much reduced speed.
This sideway expansion is limited by the inertia of the swept-up background material. 

\begin{figure}
\plottwo{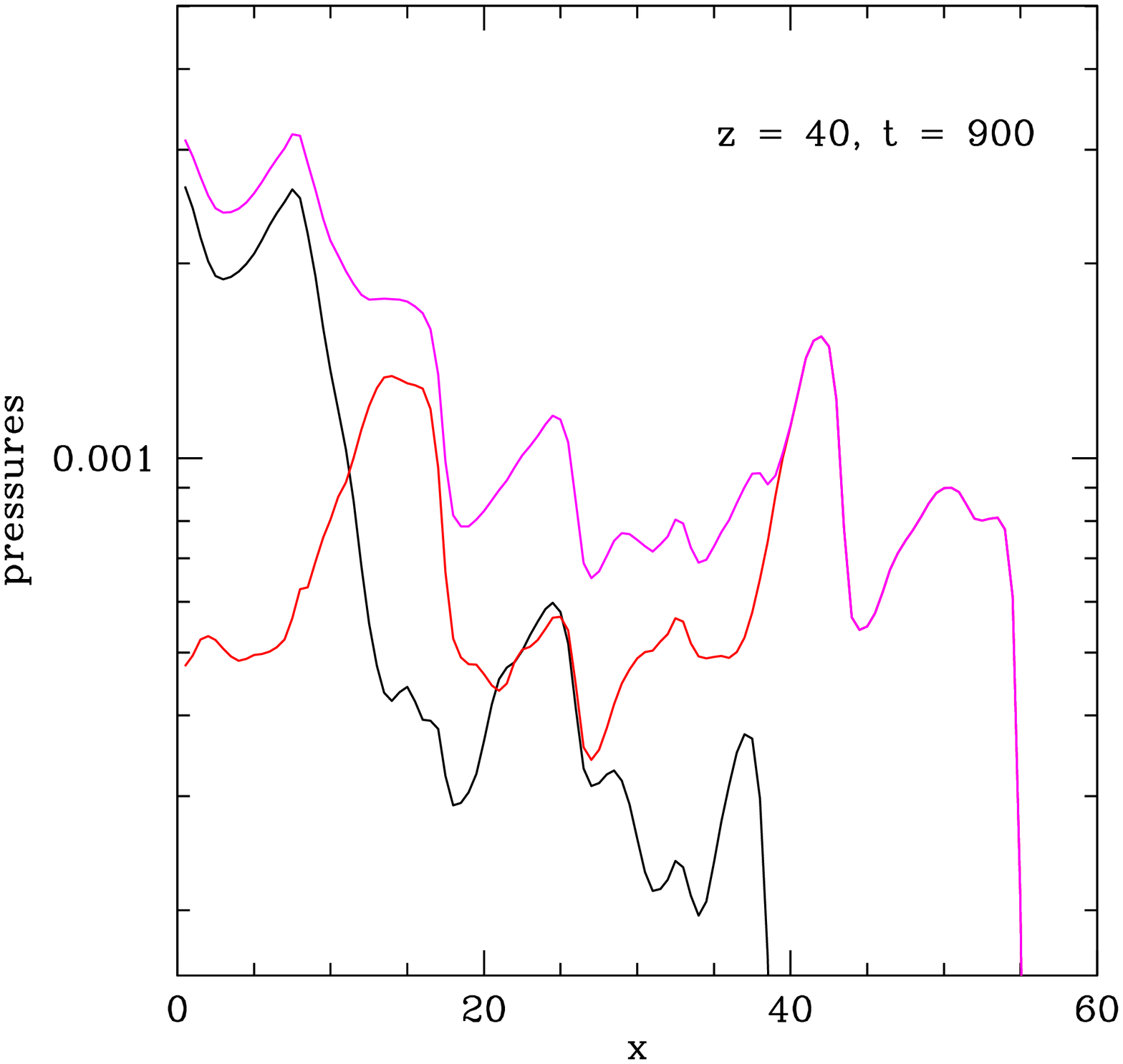}{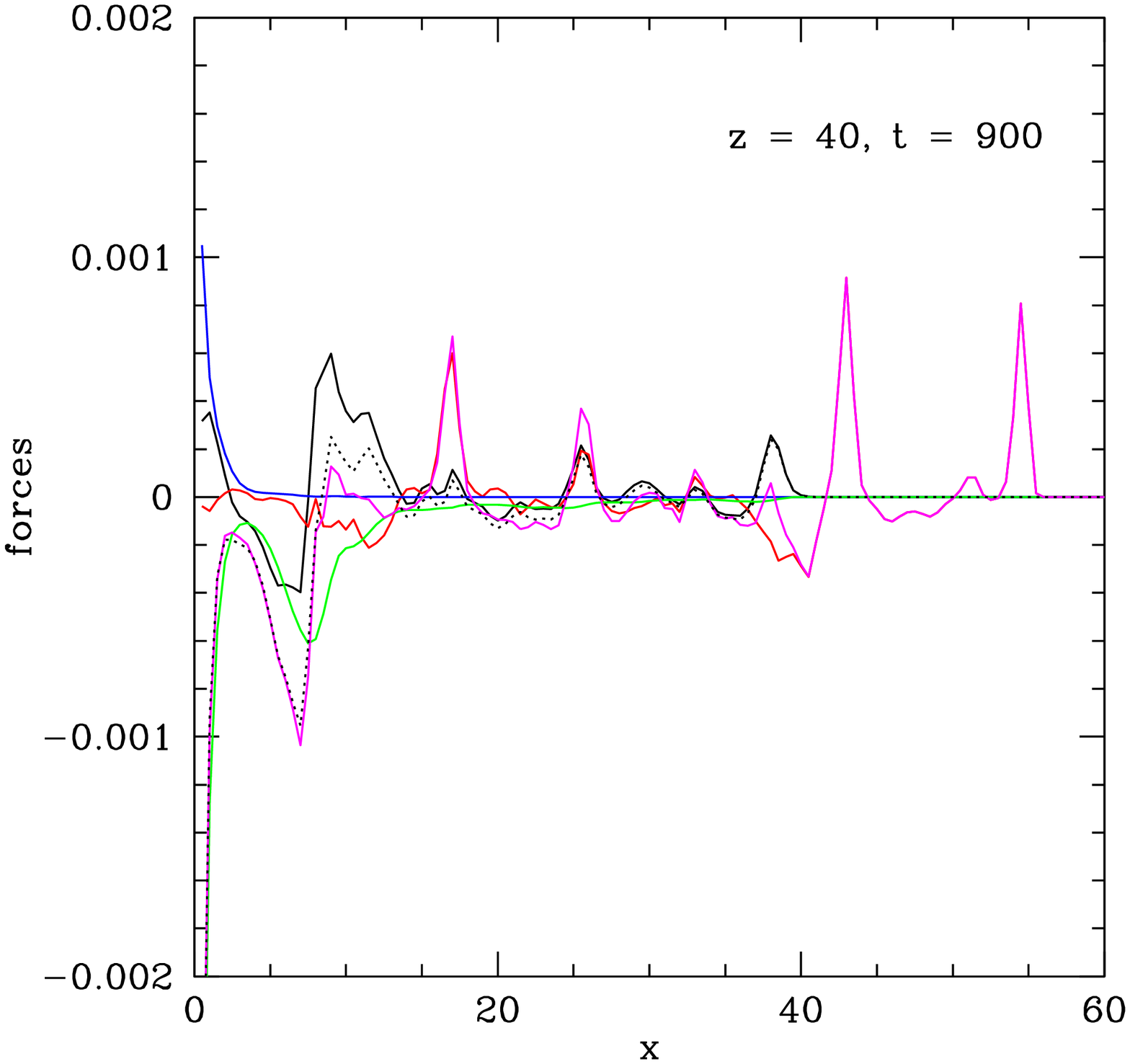}
\caption{Radial profiles of physical quantities along the x-axis in the equatorial 
plane with $(y,z) = (0, 40)$ at $t = 900$ in the fiducial run. 
Left: pressures in the radial direction. 
Black: magnetic pressure $p_{\rm m}$; 
red: gas pressure $p$; magenta: total pressure $p+p_{\rm m}$. 
Right: forces in the radial direction. 
Black solid: magnetic pressure gradient $F_{\rm mp}$; 
red: gas pressure gradient $F_{\rm p}$ ; 
blue: centrifugal force $F_{\rm c}$; green: magnetic tension force $F_{\rm t}$; 
black dotted: sum of magnetic pressure gradient and 
tension force $F_{\rm \bf J\times B} = F_{\rm mp} + F_{\rm t}$; 
magenta: total of magnetic forces and gas pressure gradient 
$F_{\rm total} = F_{\rm \bf J\times B} + F_{\rm p}$. }.
\label{fig:force480}
\end{figure}

To understand better why the magnetic structure is highly collimated along the central 
axis, we consider the force balance in the radial direction for the fiducial model, 
at $t = 900$, and vertical height $z = 40 r_{\rm inj}$, as shown in Figure  \ref{fig:force480}.
We choose $z = 40 r_{\rm inj}$ because at 
this height the jet is still quite axisymmetric, has propagated far enough in the 
vertical direction, and non-linear features from instabilities are not severe.  
At this height, the magnetically dominated part of the jet extends
from $ x = 0$ to $\sim 10 r_{\rm inj}$, with the return current located at 
$x \sim 40 r_{\rm inn}$. The outer edge of the hydrodynamic shock is located
at $x \sim 55 r_{\rm inj}$. 
The left panel of Figure \ref{fig:force480} shows that inside $x \sim 10 r_{\rm inj}$, 
magnetic pressure $p_{\rm m}$ dominates over gas pressure $p$ $(\beta \ll 1)$ 
while both keep a relative flat distribution along the radial direction; 
outside $x \sim 10r_{\rm inj}$, magnetic pressure starts to drop quickly 
while gas pressure continues to rise until $x \sim 15r_{\rm inj}$. 
We can compare this result to the analysis of non-relativistic MHD 
simulation of \cite{naka06} (their Figure 10). At large radial distances,
$x \sim 55 r_{\rm inj}$, since the plasma pressure is much larger than the 
background pressure $\sim 10^{-4}$, the radial expansion of the jet structure
is limited by plasma inertial. 

The right panel shows the various forces in the radial direction: 
near the inner jet edge, in the $10r_{\rm inj} \leq x \leq 15r_{\rm inj} $ region, 
the dominant force is the outward magnetic pressure gradient 
$F_{\rm mp} = -\del_{r}(B_{\phi}^2 +B_{z}^2)$, and there is also a smaller 
inward magnetic tension force $F_{\rm t} = -B_\phi^2/r$. The sum of the two, 
the total Lorentz force $F_{\rm \bf J\times B}$ is slightly larger than the 
inward gas pressure gradient $F_{\rm p} = -\del_{r}p$, although the magnitudes 
of the two are comparable. Inside $ x \sim 10r_{\rm inj}$, the largest force 
is the inward magnetic tension force $F_{\rm t}$ provided by the 
strong toroidal field, which gives a pinch effect. This effect is largely 
consistent with the effects of magnet hoop stress in the ``magnetic tower'' 
models \citep{lyndenbell96, lyndenbell03,li01}. There is a small rotation of gas 
that has also been produced near the axis as seen by a non-trivial 
outward centrifugal force $F_{\rm c} = \gamma \rho v_\phi^2/r$. Further out 
from the jet axis, all the magnetic forces varnish and we can see a few 
hydrodynamic shock wave fronts. It is also worth pointing out that 
although our jet is magnetically dominated (see the magenta curve, 
sum of gas pressure gradient and Lorentz force $F_{\rm total}$), 
it is not exactly force-free, as ${\bf J \times B}$ is not exactly zero 
inside the jet (black dotted curve). Furthermore, the non-zero total force
also implies that the jet is not in a force balance.

The jets we have obtained in these simulations are mildly relativistic, with 
the largest Lorentz number about $\gamma \sim 3$ (although the jet front is slowed down by the shocks), while the small amount 
of injected mass has an injection velocity of $v_{\rm inj} = 0.5c$ initially. 
Acceleration is therefore achieved through magnetic processes and 
we have observed $\gamma_{\rm max}$ increases with time 
with no signs of slowing-down. 
Due to the limit of the computational resources, we have not yet 
been able to determine the terminal speed of the jet in our models. 
However, it is plausible that a higher flux injection rate and/or 
a higher $\alpha$ can lead to a higher speed. Another issue is 
purely numerical: in RMHD/GRMHD simulations there is a 
small amount of mass loading, and the choice of density 
floor probably affects strongly $\gamma_{\rm max}$ (e.g. \cite{mg04}). 
In our simulations we have also injected a small amount of gas in 
the injection region (see discussion below), which helps us to 
maintain the validity of the RMHD integration scheme, 
especially in the injection region where the magnetic field is the strongest.

These jets also display current-driven instabilities and undergo 
subsequent strong dissipations. 
However, the jets are not disrupted and are able to propagate to large distances in 
our simulations. The cylindrical jet current is unstable most to the $m=1$ kink mode, 
which undergoes an initial period of exponential growth. Depending on the model 
parameters, outside a few tens to hundreds of $r_{\rm inj}$, the mode growth slows 
down and the non-linear interaction among the modes leads to apparent non-linear 
features  such as filaments in the current and occasional large scale ``wiggles'' in 
the jet spine. Large amounts of dissipation are also introduced outside this region. 
As a consequence, as the jet propagates further away from its launching location, 
much of magnetic energy has been transformed into jet kinetic energy and heat, 
although the jet is still collimated and continues to propagate, albeit at a slower speed. 
We notice that although the $m = 1$ mode grows exponentially, its power remains smaller 
than the power in the $m = 0$ mode throughout the simulations. 
This is consistent with the fact that the jet is not disrupted even with CDI present. 
Such non-disruption behavior of jet is consistent with the past RMHD simulations. 
These results also support the idea some other mechanisms may be at work to 
suppress the non-linear impact of CDIs (e.g., \cite{narayan09}). 

We suggest that the ability of jets to avoid the complete disruption is due to both
the rapid jet propagation and the fast evolution of the associated underlying 
magnetic structure, which we collectively term ``dynamic stabilization''. 
Away from the injection region, the Alfv\'{e}n speed in the magnetized region
decreases from $\sim 0.9c$ near the central spine to $\sim 0.2-0.6c$ near the boundaries. 
The background flow (except that near the jet front), however, still has a relativistic speed 
of $>0.9c$.  It is therefore possible that this fast background flow has 
modified the physical quantities faster than the instability growth timescale. 
The same arguments can be applied to the large $\alpha$ runs when the 
magnetic structure tends to evolve even faster.  
In other words, the CDIs developed in our simulated jets are quite convective, 
rather than being absolute instability. 
To the extent we can simulate
the jet propagation, it remains collimated and propagating at a steady speed. 
It therefore remains to be seen how dynamical stabilization will continue to 
help jets survive the instabilities and whether the environmental factors may
play some additional roles in determining the fate of relativistic jets.  

We have also shown that as these current-carrying jets propagate far from 
the injection region, magnetic energy can be transformed into kinetic 
energy of the jet and also generates heat. The magnetization 
parameter $\sigma$, although much larger than one at the jet base,  
can become much smaller with $\sigma \ll 1$ in the region 
where CDIs have grown to display non-linear features. 
Note that in our model the smaller $\sigma$ is not a result of the jet 
shocking on the external medium, but a consequence of 
development of CDIs in a current-carrying jet. Although many 
non-linear features of CDIs appear in our models, the model 
has not reached a saturated state: all the energetics in the 
models still increase over time and it is not clear what the 
jet dynamics will be on an even longer time scale. Future 
simulations of larger computational domain with longer evolution time 
are needed to give a more comprehensive picture of the 
$\sigma$ question.  Recent local simulations of CDIs by \cite{oneill12} 
have also shown that development of CDIs are able to convert 
magnetic energy into kinetic energy and thermal energy, and 
they also have not found a saturated state. Nevertheless, all 
these simulations are starting to show that CDIs are indeed 
able to tackle the $\sigma$ problem. 

In our high resolution run we have observed some large scale wiggles 
near the jet fronts\footnote{These wiggles are also seen in model with 
a thicker disk, with the same resolution with the fiducial run, 
although not shown in the paper}.  In the future it would be 
interesting to see whether these models are able to produce 
knots and spots along the jet axis, which are often observed in 
AGN jets.  All our models also produce a central current (``spine'') 
along the vertical axis, and a cocoon-like return current which 
locates at a large distance from the jet axis and encloses 
the central jet. In the high resolution run this return current 
also exhibits non-axisymmetric features. These return currents 
have also been produced in the past MHD simulations 
(e.g. \cite{ukrl06, li06, naka06, naka07, naka08}). It would be 
interesting to see whether these large scale return currents are 
observable (e.g. \cite{kron11}). Time-dependent jet properties 
produced in this work, when combined with radiative processes, 
can also be used to compare with observational features of 
AGN jets, such as their time variability\footnote{The time resolution 
of our simulation is on the order of days.}. This work marks 
our first effort toward producing AGN jet diagnostics from a 
numerical RMHD model.  

It would be useful to scale the model parameters for a supermassive 
black hole system. As discussed in \S 2, for a 
$3\times 10^9 \msun$ black hole as the one at the center of M87, 
we have a magnetic energy injection rate of $5.2\times 10^{46}\ergs$ 
(the Eddington luminosity is $L_{\rm Edd} \sim 3.9\times10^{47}\ergs$). 
This current-carrying jet can propagate from its injection region of size 
$r_{\rm inj} \sim 0.14\pc$ to a distance of $\sim 28\pc$ in the fiducial model, 
and to a distance of $\sim 56$ pc in the $\alpha =40$ model,  
without being disrupted. The features of CDIs show up on the pc scales. 
The magnetic field has a strength that is on the order of $10^{-3}{\rm G}$ 
in the jet axis and far from the core. The total current is estimated to be 
$I \sim 10^{18}{\rm amp}$ in the fiducial model. For the background gas, 
we have adopted a uniform background density of $10^2 {\rm cm}^{-3}$ 
and temperature of $5 {\rm keV}$.  We will explore the effect of 
background profile in the future investigation. We have also injected 
a small amount of gas in the injection region, and in the fiducial 
model the mass injection rate 
is $\dot{M}_{\rm inj} \sim 0.09\msun{\rm yr^{-1}}$, which is 
much smaller than the Eddington accretion rate $\dot{M}_{\rm Edd} 
\sim 13 \msun{\rm yr^{-1}}$. 
(Usually we need to inject more mass if the magnetic energy injection rate 
is increased due to numerical  reasons.)
Lastly, for the resolution, in the fiducial model the smallest length scale 
is $\Delta l \sim 0.01\pc$ and the smallest time scale is $\Delta t \sim 20$ days. 
Note this time scale is still long compared to the time scale on which the 
TeV flares operate. Therefore, pushing to higher resolution deserves 
more efforts in future studies.  

We have investigated the fiducial model with two different 
resolutions, and both exhibit qualitatively similar behaviors. 
However, the convergence is not achieved: this is especially true for 
the instability and the shocks; effect of resolution on energy transition 
is not clear yet. We will leave the even higher resolution studies to 
the future work.  We have also investigated a model with a higher 
toroidal-to-poloidal injection ratio. The details of the injection function 
definitely affect jet properties. In the future, we will explore more 
model parameters including magnetic field geometries, injection functions, 
and external environment profiles 
(e.g. power-law external pressure profiles used in \cite{komi09}).

We have chosen an injection model that has closed poloidal field lines, 
which causes $B_z$ change directions beyond $r_{\rm inj}$ with no net-flux.  
Different field injection configuration exists. For example, past GRMHD 
black hole accretion simulations have explored models with initial 
configurations with open field lines/net flux  (e.g. "Magnetically Arrested Disc" models). 
However, whether the disk has net-flux or not is a un-resolved question, 
largely owing to our lack of knowledge of disk dynamo.  Since 
these past simulations have not typically produced the jet structure at large scales 
where comparison with observations becomes more feasible,  it is therefore of 
interest to explore the case with zero net-flux.  Furthermore, studies of 
large-scale jets in the intra-cluster medium (hereafter ICM;  \cite{li06,naka06,naka07,naka08}) 
have argued that magnetic tower model provides good fits to observations 
of jetÕs morphologies in the ICM.  Future work is therefore needed to explore 
different initial field configurations and their consequences in jet stability and dissipation.

Lastly, we want to point out that recently there has been great progress in 
the laboratory experiments to study current-driven instabilities in jets 
(e.g. \cite{bellan05,ber06}).  Although the physical conditions in our AGN jet 
models differ greatly from the parameters in laboratory jets (e.g. density, current etc.), 
it would be of great interest to see whether laboratory plasma experiments can 
teach us the general principles in understanding astrophysical jets.
  
\acknowledgements
The authors are grateful to Stirling Colgate, Brenda Dingus, Ken Fowler, 
John Hawley, Philipp Kronberg, and Masanori Nakamura for discussions.  We also thank the anonymous referee for insightful suggestions. This work is  supported by the LDRD and Institutional Computing Programs at LANL and by DOE/Office of Fusion Energy Science through CMSO.

\newpage 
\begin{deluxetable}{llll}
\setlength{\tabcolsep}{0.07in}
\tabletypesize{\scriptsize}
\tablecolumns{4}
\tablecaption{Units of Physical Quantities For Normalization \label{units}}
\tablewidth{0pt}
\tablehead{
\colhead{Physical Quantities} & \colhead{Description} & \colhead{Normalization Units} & \colhead{Typical Values} \\
}
\startdata
$r$ $[=(x^2+y^2+z^2)^{1/2}]$ & Length & $r_{\rm inj}$ &   0.143pc\\
$v$                & Velocity field & c & $3.0\times 10^{10}\cms$ \\
$t$                & Time & $r_{\rm inj}/c$ & 0.47yr  \\
$\rho$            & Density & $\rho_0$ & $1.67\times 10^{-22}\gcm$  \\
$P$               & Pressure & $\rho_0c^2$ & $0.15{\rm dyncm}^{-2}$ \\
$B$               & Magnetic field & $(4\pi \rho_0 c^2)^{1/2}$ & $1.38{\rm G}$\\
$E$               & Energy    & $B^2 r_{\rm inj}^3/(8\pi)$ & $6.52\times10^{51}\erg$\\
$P$               & Power     & $B^2 r_{\rm inj}^2c/(8\pi)$ & $4.42\times10^{44}\ergs$\\                         
\enddata
\end{deluxetable}

\clearpage

\clearpage


\clearpage

\clearpage

\clearpage

\clearpage


\clearpage

\clearpage
\end{document}